\documentclass[journal]{new-aiaa}
\usepackage[utf8]{inputenc}
\usepackage{textcomp}
\usepackage{algorithm}
\usepackage{tabularray}
\usepackage{graphicx}
\usepackage{subcaption}
\usepackage{amsmath}
\usepackage[version=4]{mhchem}
\usepackage{siunitx}
\usepackage{multirow}
\usepackage{longtable,tabularx}
\newtheorem{remark}{Remark}

\newtheorem{assumption}{Assumption}
\DeclareMathOperator*{\minimize}{minimize}

\title{Model Predictive Guidance for Fuel-Optimal Landing of Reusable Launch Vehicles}
\author{Ki-Wook Jung\footnote{Ph.D. Candidate, Department of Aerospace Engineering; jkw1031@kaist.ac.kr}, Sang-Don Lee\footnote{Ph.D. Candidate, Department of Aerospace Engineering; sedan96@kaist.ac.kr}, Cheol-Goo Jung\footnote{Ph.D. Candidate, Department of Aerospace Engineering; jjulgoo22@kaist.ac.kr} and Chang-Hun Lee\footnote{Associate Professor, Department of Aerospace Engineering; lckdgns@kaist.ac.kr}}
\affil{Korea Advanced Institute of Science and Technology, Daejeon 34141, Republic of Korea}

\begin{document}

\maketitle

\begin{abstract}
This paper introduces a landing guidance strategy for reusable launch vehicles (RLVs) using a model predictive approach based on sequential convex programming (SCP). The proposed approach devises two distinct optimal control problems (OCPs): planning a fuel-optimal landing trajectory that accommodates practical path constraints specific to RLVs, and determining real-time optimal tracking commands. This dual optimization strategy allows for reduced computational load through adjustable prediction horizon lengths in the tracking task, achieving near closed-loop performance. Enhancements in model fidelity for the tracking task are achieved through an alternative rotational dynamics representation, enabling a more stable numerical solution of the OCP and accounting for vehicle transient dynamics. Furthermore, modifications of aerodynamic force in both planning and tracking phases are proposed, tailored for thrust-vector-controlled RLVs, to reduce the fidelity gap without adding computational complexity. Extensive 6-DOF simulation experiments validate the effectiveness and improved guidance performance of the proposed algorithm.
\end{abstract}

\section*{Nomenclature}

\noindent(Nomenclature entries should have the units identified)

{\renewcommand\arraystretch{1.0}
\noindent\begin{longtable*}{@{}l @{\quad=\quad} l@{}}
$C_L$  & lift coefficient \\
$C_D $ & drag coefficient \\
$C_z $ & lateral aerodynamic force coefficient \\
$l_{cp}$& relative distance from center of gravity to center of pressure\\
$l_{c}$& relative distance from center of gravity to TVC hinge\\
$s_{ref}$& reference area of the rocket \\ 
$d_{ref}$& reference length of the rocket \\
$\boldsymbol{r}$ & position vector of vehicle, $\in \mathbb{R}^{3 \times 1}$ \\
$\boldsymbol{v}$ & velocity vector of vehicle, $\in \mathbb{R}^{3 \times 1}$ \\
$\boldsymbol{T}$ & thrust vector of vehicle, $\in \mathbb{R}^{3 \times 1}$ \\
$\boldsymbol{F}_{aero}$ & aerodynamic force, $\in \mathbb{R}^{3\times1}$ \\
$\boldsymbol{F}_{drag}$ & aerodynamic drag force, $\in \mathbb{R}^{3\times1}$ \\
$\boldsymbol{F}_{lift}$ & aerodynamic lift force, $\in \mathbb{R}^{3\times1}$ \\
$\boldsymbol{g} $ & gravitational acceleration of the Earth, $\in \mathbb{R}^{3\times1}$\\
$\alpha_T$ & total angle of attack \\
$g_{ref}$ & reference gravitational acceleration of the Earth \\
$I_{sp}$ & specific impulse of the rocket\\
$\rho$ & density of the air\\
$\bar q$ & dynamic pressure, $1/2 \rho \lVert \boldsymbol{v} \rVert^2 $\\ 
\multicolumn{2}{@{}l}{Subscripts}\\
$\mathcal{I}$ & inertial frame \\
$\mathcal{B}$ & vehicle body frame \\
$k$ & parameter at the $k$-th time node \\ 

\multicolumn{2}{@{}l}{Superscript}\\
$'$ & time derivative in normalized time $\tau$ \\
$i$ & optimal solution at the $i$-th iteration \\
$p$ & parameters related to the planning problem \\
$t$ & parameters related to the tracking problem \\
\end{longtable*}}

\section{Introduction}
In recent years, reusability has emerged as one of the key paradigms in launch vehicle development. A reusable launch vehicle (RLV) has the capability to retrieve and reuse part of the vehicle after the launch mission. If the stable operation of the RLV fleet is feasible, which is already demonstrated by the private sector~\cite{reddy2018}, the benefits can range from improving operational flexibility to lowering the economic hurdles to reach space. Existing guidance and control schemes for atmospheric reentry and landing have been developed and verified by programs such as NASA's Space Shuttle~\cite{harpold1983, kafer1982}. However, the advent of reusable launchers, particularly the return of first-stage boosters, necessitates new guidance algorithms for inexperienced flight configurations. Among the various flight phases of the RLV, the most challenging part for the guidance algorithms is widely deemed as the landing burn~\cite{blackmore2016}, since the vehicle has to autonomously decide the landing trajectory that satisfies optimality, safety, and accuracy using limited control method: thrust vector control (TVC) with bounded throttling capability. 

Retro-propulsion for planetary landing, where the thrust of a rocket engine counters the velocity vector to decelerate, was pioneered during the space race era in the 1960s. Analytic landing guidance strategies that compute the acceleration commands based on the current state and time-to-go were developed and used for the Apollo landing module~\cite{cherry1964, Klumpp1974}. As exploration missions become more advanced, considerations such as fuel consumption optimality, thrust magnitude limits, and various path constraints have become essential. Although challenging, deriving an analytical solution is not entirely impossible, as several works~\cite{Guo2013, Wang2021two} have been developed to account for certain path constraints and thrust limits using the feedback guidance in~\cite{cherry1964}. These analytic methods are still attractive since they can provide closed-loop performance in nature. However, as requirements vary, deriving such a solution becomes increasingly challenging and often requires numerical approaches. 

Alongside the progress of computational power in on-board computers, computational guidance methodologies, such as trajectory optimization techniques that solve the optimal control problem (OCP) directly via parameter optimization, have gained attention. Among the various parameter optimization algorithms, convex programming (CP) possesses desirable properties for on-board applications, as it guarantees convergence within a bounded number of iterations~\cite{boyd2004convex}. Açıkmeşe and Ploen~\cite{Acikmese2007} proposed solving a single second-order cone programming (SOCP) problem to generate powered-descent trajectories while considering various state constraints and thrust bounds, which opened the door to using CP in aerospace applications. This line of research extended to the minimum landing error problem~\cite{Blackmore2010}, and was verified by numerous flight tests with dedicated vertical take-off vertical landing (VTVL) test vehicles and on-board algorithms~\cite{Scharf2017, Dueri2017}, marking an important milestone demonstrating the feasibility and practicality of CP in powered-descent guidance (PDG). 

Although SOCP can handle a relatively wide range of OCPs, it is still confined to problems with linear dynamics and inequalities defined under positive or second-order cone (SOC). As the pursuit of improving guidance performance by bridging the gap in model fidelity between OCP and the actual system continued, the limitations of using pure CP became evident, especially in aerospace applications where highly nonlinear dynamics and non-convex path constraints are inevitable. Therefore, many works focused on methods to expand the envelope of CP using a sequential approach~\cite{liu2014, Mao2016, SzmukAIAA, wang2017}, namely successive linearization or sequential convex programming (SCP). This expansion enabled the solution of a non-convex 3-DOF problem with quadratic drag through successive linearization in~\cite{SzmukAIAA}. More accurate representations of aerodynamic drag and lift forces were handled in~\cite{XinfuExact, WangMPC, Jung2023}, with convex relaxation approaches proposed to reduce the non-convexity of OCPs in~\cite{XinfuExact, Jung2023}. Another aspect of enhancing fidelity is the inclusion of rotational dynamics, which requires a 6-DOF representation of the RLV dynamics~\cite{szmuk2017, Szmuk6DSTC, sagliano2021}. Szmuk et al. proposed an ellipsoidal aerodynamics model with rigid-body rotational motion in~\cite{Szmuk6DSTC}. Conversely, accurate aerodynamic force coefficients were used in~\cite{sagliano2021}, but with simplified rotational dynamics, where each attitude is modeled as a first-order linear system. 

While the technique of obtaining optimal trajectories via SCP has vastly improved, a few degrees of freedom still exist in designing a landing guidance algorithm that leverages such developments. A classical way of utilizing the optimal reference trajectory is to design an analytical tracking law that tracks the given trajectory~\cite{bollino2006, LuPDG2}. This scheme has also been applied in flight tests in~\cite{Scharf2017}. More recently, an integrated tracking guidance and control algorithm using modern control theory such as H-infinity control was proposed for the aerodynamic descent phase of the RLV in~\cite{Sagliano2023a}. Similar to the analytical guidance algorithms, these tracking algorithms have performance limitations imposed by system dynamics, which are inherently slow due to the structural constraints of launch vehicles~\cite{Orr2014}, and cannot directly consider complex constraints. Instead, adequate safety margins should be incorporated at the trajectory design stage, while fully leveraging such margins through the tracking law can also pose challenges. If the trajectory optimization can be executed swiftly in real-time, feed-forwarding the optimal control sequence from the trajectory can be another promising methodology to design a landing guidance in a model predictive context. Wang et al. proposed a model predictive approach by formulating a recursively feasible optimal control problem~\cite{WangMPC}. This method ensures a stable stream of optimized thrust, but the solution update was relatively slow, as it required computing the full trajectory, while showing high landing velocity under aerodynamics uncertainties. Utilizing a formulation similar to that in~\cite{Szmuk6DSTC}, the study in~\cite{Guadagnini2022} assessed landing guidance performance by recursively solving a 6-DOF fuel-optimal problem, then feed-forwarding the resulting thrust vector into the control loop. However, handling the complex OCP often became infeasible with model uncertainties, and the computational effort required for the 6-DOF problem is substantial, leading to delayed solution updates. This lag in determining guidance commands may degrade guidance performance with diminishing closed-loop property.

Therefore, concurrently achieving model fidelity and computational efficiency is crucial but challenging in the development of landing guidance through trajectory optimization. Given that on-board flight processors are significantly slower than modern desktop processors, reducing the computational cost is further important for realizing computational guidance schemes. Many previous works have focused on developing customized convex solvers~\cite{Dueri2017, pei2023, kamath2023} that leverage the specific properties of each OCP, greatly reducing the computation time. Nevertheless, the complexity of formulating sophisticated problems renders fast solving of corresponding OCPs on-board increasingly difficult, posing a significant barrier to adopting model predictive approaches in on-board applications. This trade-off relationship foreshadows that a balanced formulation between model fidelity and subsequent computational effort may lead to an effective landing guidance algorithm that utilizes model predictive methodologies.

Based on these observations, this paper proposes a landing guidance scheme for an RLV, which incorporates the model predictive approach and trajectory optimization via the SCP algorithm. To address challenges encountered in implementing numerical optimization methods, we propose a guidance scheme composed of two parts: planning and tracking tasks. In the planning task, we obtain a fuel-optimal landing trajectory through full prediction to the landing site based on SCP, while considering the practical constraints inherent to RLVs. Instead of employing an analytical tracking guidance law, we determine the optimal tracking command through SCP, following a methodology similar to the receding horizon concept in model predictive control (MPC)~\cite{kwon2005}. By reducing computational load with a tunable prediction horizon length, we aim to secure a relatively fast solution update, thereby harnessing near closed-loop performance. We broaden the feasible region for state and control variables in the tracking task beyond the planning phase, aiming to improve both the feasibility of the problem and robustness to uncertainties. To improve the model fidelity of the tracking task, we propose integrating rotational dynamics in a simplified representation that enables an input-affine form of the OCP. This results in guidance commands that compensate for the vehicle's transient response and improves the numerical stability of solving the OCP~\cite{XinfuExact}. Moreover, to account for thrust vectoring induced by the attitude control loop, we propose modifications to the aerodynamic force representation for both the planning and tracking tasks. Extensive numerical experiments in 6-DOF simulation are conducted to verify the effectiveness and overall performance of the proposed guidance algorithm.

The contributions of this study are threefold. Firstly, it demonstrates the concurrent use of trajectory optimization in both trajectory planning and tracking, harnessing the benefits of numerical methods while mitigating their inherent challenges through the tracking task. Secondly, it introduces a novel approach for accounting for the control loop characteristics of RLVs by modifying the aerodynamic force representation. This adjustment is tailored for thrust-vector-controlled RLVs and capitalizes on their characteristics. The modification significantly narrows the model fidelity gap that arises from the simplification of the rotational dynamics, without imposing extra computational burden or complexity, thereby substantially enhancing guidance performance. Lastly, a new problem formulation for landing guidance is proposed, incorporating significant practical path constraints and decision variables such as the aerodynamic load limit, which have not been considered in previous works. This enhancement improves the practicality of the landing guidance algorithm.

The remainder of this paper is structured as follows: Section~\ref{sec2} outlines the two OCPs for the planning and tracking tasks. Section~\ref{sec3} details the SCP algorithm, the primary method for solving non-convex problems. In Section~\ref{sec4}, the features and characteristics of the proposed guidance algorithm are discussed, with its evaluation and validation through numerical studies presented in Section~\ref{sec5}. Finally, concluding remarks are offered in Section~\ref{sec6}.

\section{Problem Formulation}\label{sec2}
This section outlines the formulation of the OCPs for the proposed guidance algorithm. The proposed method is divided into two primary functions: the planning and tracking of the trajectory, each necessitating a distinct definition of OCPs. 

\subsection{Planning Problem}
In the planning problem, a complete trajectory from the aerodynamic descent to the landing at the desired position is included to provide an optimal trajectory that minimizes fuel consumption, including engine ignition time. Hereafter, the planning problem is referred to as Problem $\mathcal{P}$.

\subsubsection{Vehicle Dynamics}
\begin{figure}[!htb]
    \centering
    \includegraphics[width=7cm]{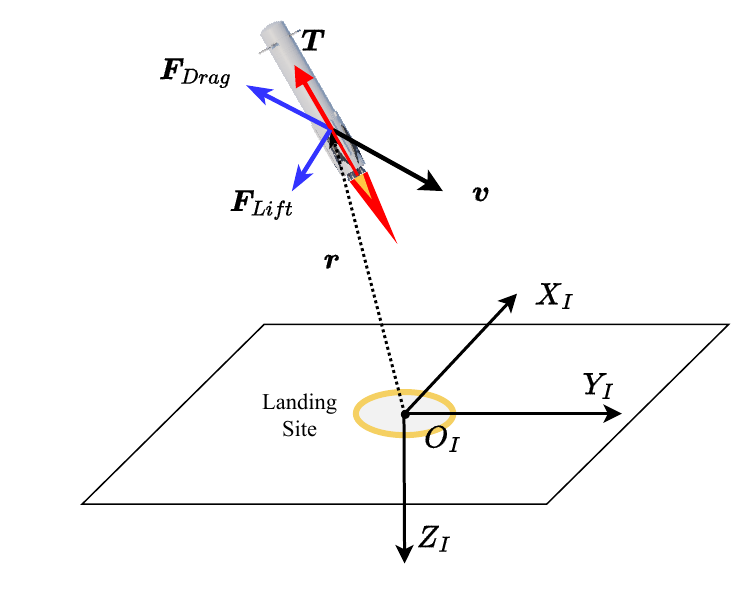}
    \caption{Reference frame and vectors.}
    \label{fig:kine}
\end{figure}

First, the state and control variables of Problem $\mathcal{P}$ are defined as follows:
\begin{equation} \label{eq:pxu}
    \begin{gathered}
        \boldsymbol{X} = \left[ \boldsymbol{r}, \boldsymbol{v}, m \right]^T, \quad \boldsymbol{U} = \boldsymbol{T},\quad \boldsymbol{Z} = \left[ X, U \right]^T.
    \end{gathered}
\end{equation}
where $\boldsymbol{r}, \boldsymbol{v}$, and $\boldsymbol{T}$ represent position, velocity, and thrust vector, respectively, while $m$ is the mass of the vehicle. Given that a typical RLV has a relatively short duration of the landing burn with several kilometers change in altitude, the following assumption is established.
\begin{assumption}
    The Earth is considered flat and non-rotating, with a constant gravitational acceleration.
    \label{asm:dynamics}
\end{assumption}
Based on Assumption~\ref{asm:dynamics}, a North-East-Downward (NED) frame with the origin residing at the desired landing position is used as the reference frame for the proposed guidance, referred to as the inertial frame ($\mathcal{I}$-frame). Figure~\ref{fig:kine} depicts the landing scenario and vector definitions for the RLV, with the subscript ``$\mathcal{I}$'' denoting the $\mathcal{I}$-frame. 
To ease the burden of computational complexity and enable rapid optimal trajectory generation, the 3-DOF point mass model is adopted in the formulation of the OCP. This lack of model fidelity in the planning problem is later compensated for by modified lift force and the proposed tracking phase. The vehicle dynamics is given as follows.
\begin{equation} \label{eq:pdyn}
    \begin{aligned}
    \dot {\boldsymbol{r}}  &=   \boldsymbol{v} \\
    \dot {\boldsymbol{v}}  &=  \frac{\boldsymbol{T} + \boldsymbol{F}_{aero}}{m} + \boldsymbol{g} \\
    \dot m  &=  - \frac{\lVert \boldsymbol{T} \rVert }{g_{ref}I_{sp}} - \frac{P_e A_{exit}}{g_{ref}I_{sp}}\\
    \end{aligned}
\end{equation}
where $\boldsymbol{g} = \left[ 0, 0, g_{ref} \right]^T$. For the purpose of brevity, the vector expressed in the $\mathcal{I}$-frame is written without subscript. The operator $\lVert \cdot \ \rVert$ denotes the $L_2$ norm, $P_e$ is the atmospheric pressure, and $A_{exit}$ is the total exit area of the rocket nozzle. The term $P_e A_{exit}$ is added to the equation for the depletion of mass to reflect the back pressure loss of the rocket engine.

The high dynamic pressure environment further emphasizes the importance of accurately modeling aerodynamic forces in generating realistic reference trajectories. In this work, the lift and drag forces constitute the aerodynamic force $\boldsymbol{F}_{aero}$ as follows:
\begin{equation} \label{eq:aero1a}
    \boldsymbol{F}_{aero}   =   \boldsymbol{F}_{drag} + \boldsymbol{F}_{lift} 
\end{equation}
with 
\begin{equation} \label{eq:aero1}
    \boldsymbol{F}_{drag}   =   -\bar q s_{ref} C_D\frac{ \boldsymbol{v}}{\lVert \boldsymbol v \lVert }, \quad
    \boldsymbol{F}_{lift}   =   \bar q s_{ref} C_L \hat {\boldsymbol{L}}
\end{equation}
where $\hat {\boldsymbol{L}} =  \frac{\left(\boldsymbol{T}\times \boldsymbol{v}\right)\times \boldsymbol{v}}{\lVert {\boldsymbol{T} \times \boldsymbol{v}}\rVert \lVert \boldsymbol{v} \rVert}$. In the derivation of the unit vector $\hat {\boldsymbol{L}}$ that represents the direction of the lift force, the following assumptions are utilized. 
\begin{assumption}
    The vehicle is symmetric about the roll axis.
    \label{asm:rollsym}
\end{assumption}
\begin{assumption}
    The longitudinal axis of the vehicle is aligned with the thrust direction.
    \label{asm:thrustalign}
\end{assumption}
The RLV usually has an asymmetric shape in the roll axis due to the existence of aerodynamic fins. Despite this, Assumption~\ref{asm:rollsym} remains valid, as the lift generated by the aerodynamic fins is significantly smaller than that generated by the rocket body. Assumption~\ref{asm:thrustalign} is widely used for various 3-DOF PDG problems~\cite{Acikmese2007, Scharf2017, LuPDG}. Based on Assumption~\ref{asm:rollsym} and Assumption~\ref{asm:thrustalign}, the lift force is perpendicular to the velocity vector, $\boldsymbol{v}$, and lies on the plane spanned by $\boldsymbol{T}$ and $\boldsymbol{v}$, which is referred to as plane $\mathcal{A}$ hereafter. To simplify the mathematical expression of $\boldsymbol{F}_{lift}$ that involves a fraction with vector product and norm, the following assumption is utilized.
\begin{assumption}
    The angle of attack experienced by the vehicle during the landing burn is moderately small.
    \label{asm:aoa}
\end{assumption}
Assumption~\ref{asm:aoa} is based on the observation that an RLV's angle of attack during the high dynamic pressure environment should be limited due to the structural limits and trim envelope. Then, the lift coefficient $C_L$ can be approximated by a first-order Taylor expansion.
\begin{equation} \label{eq:cla}
    \begin{gathered}
        C_L \approx \left. \frac{\partial C_L}{\alpha_T} \right|_{\alpha_T = 0 } \alpha_T = C_{L,\alpha} \alpha_T
    \end{gathered}
\end{equation}
For a non-negative and small value of $\alpha_T$, it can be approximately expressed in terms of the thrust and velocity vector as:
\begin{equation} \label{eq:alpvec}
    \begin{gathered}
        \sin{\alpha_T} =\frac{\lVert \boldsymbol{T} \times \boldsymbol{v} \rVert }{\lVert \boldsymbol{v} \rVert  \lVert \boldsymbol{T} \rVert }  \approx \alpha_T
    \end{gathered}
\end{equation}
It should be noted that the angle of attack can have a large value and exceed $\pi/2$, especially when the dynamic pressure is low. In this case, the impact of the aerodynamic forces is relatively small, which makes the error from the approximation negligible. By substituting Eqs.~\eqref{eq:cla} and~\eqref{eq:alpvec} to $\boldsymbol{F}_{lift}$ in Eq.~\eqref{eq:aero1}, we have:
\begin{equation} \label{eq:plift2}
    \begin{gathered}
         \boldsymbol{F}_{lift}   =   \bar q s_{ref} C_L \hat {\boldsymbol{L}} = \bar q s_{ref} C_{L,\alpha} \frac{\left(\boldsymbol{T}\times \boldsymbol{v}\right)\times \boldsymbol{v}}{\lVert T \rVert }
    \end{gathered}
\end{equation}
The result in Eq.~\eqref{eq:plift2} still requires a sequence of vector products and division by the vector norm. Nevertheless, the complexity of the expression is greatly reduced, which will consequently be beneficial for computationally obtaining the solution of Problem $\mathcal{P}$.

\subsubsection{Constraints}
\begin{figure}[!htb]
    \centering
    \includegraphics[width=7cm]{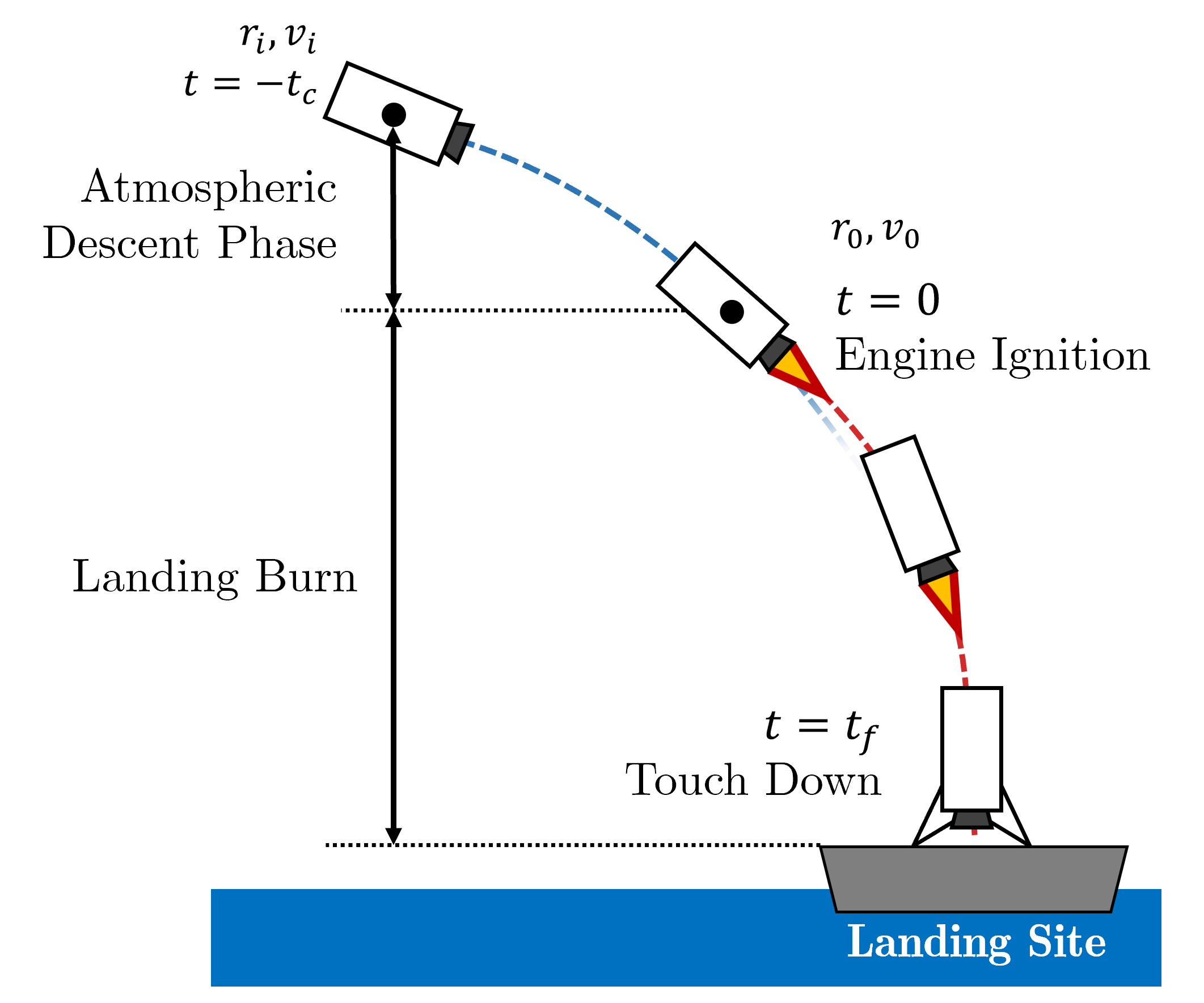}
    \caption{Flight phases of the RLV.}
    \label{fig:fphase}
\end{figure}
The RLV experiences the aerodynamic descent prior to the ignition of the landing burn, as shown in Fig.~\ref{fig:fphase}. The descent trajectory can easily be computed by numerical integration from the current state assuming zero angle of attack. Similar to previous work~\cite{Szmuk6DSTC}, the least squares fit by a quadratic polynomial of ignition time $t_c$ can accurately model the predicted trajectory as a function of $t_c$. Then, we impose the initial condition as follows to determine the optimal ignition time without introducing an additional phase in Problem $\mathcal{P}$, thereby reducing the computational burden. 
\begin{equation} \label{eq:pinit}
    \begin{aligned}
\boldsymbol{r}\left(0\right) &=  \boldsymbol{K}_r \left[t_c^2 , t_c , 1 \right]^T \\
\boldsymbol{v}\left(0\right) &=  \boldsymbol{K}_v \left[t_c^2 , t_c , 1 \right]^T \\
m\left(0\right)  &=   m_0 
    \end{aligned}
\end{equation}
where $\boldsymbol{K}_r,  \boldsymbol{K}_v \in \mathbb{R}^{3\times3}$. In Eq.~\eqref{eq:pinit}, the coefficients $\boldsymbol{K}_r$ and $\boldsymbol{K}_v$ are the result of the least squares fit of the descent trajectory. The terminal conditions are given for position and velocity to ensure a soft and precise landing at the desired position.
\begin{equation} \label{eq:pterm}
    \begin{gathered}
        \boldsymbol{r}\left(t_f\right) = \boldsymbol{v}\left(t_f\right) =  \boldsymbol{0}_{3\times1}
    \end{gathered}
\end{equation}
To prevent structural failure and violation of the trim envelope, an upper bound on the aerodynamic loads is imposed.
\begin{equation} \label{eq:paload}
    \begin{gathered}
        \bar q \alpha_T  \leq {L}_{lim}^p
    \end{gathered}
\end{equation}
where the parameter $L_{lim}^p$ is the upper bound of the aerodynamic loads for Problem $\mathcal{P}$. We can rewrite Eq.~\eqref{eq:paload} into a nonlinear function $g_L$ by using the expression of $\cos \alpha_T$ in terms of velocity and thrust vector as in Eq.~\eqref{eq:alpt}.
\begin{equation} \label{eq:alpt}
    \begin{gathered}
        \alpha_T    =   \cos^{-1} \left( -\frac{\boldsymbol{T} \cdot \boldsymbol{v}}{\lVert \boldsymbol{T}\rVert \lVert \boldsymbol{v}\rVert  }\right)
    \end{gathered}
\end{equation}
\begin{equation} \label{eq:paload2}
    \begin{gathered}
          \cos \alpha_T \geq \cos \left( \frac{L_{lim}^p}{\bar q} \right) \rightarrow
         g_L\left(\boldsymbol{X}, \boldsymbol{U}\right) \triangleq   {\boldsymbol{T} \cdot \boldsymbol{v}} + {\lVert \boldsymbol{T}\rVert \lVert \boldsymbol{v}\rVert  } \cos  \left( \frac{L_{lim}^p}{\bar q} \right) \leq 0
    \end{gathered}
\end{equation}
One advantage of the expression in Eq. \eqref{eq:paload2} is that total angle of attack becomes unconstrained when $\bar q$ is low by fixing the maximum value of ${L_{lim}^p}/{\bar q}$ as $\pi$. Therefore, this allows  for versatile maneuvering of the RLV when the effect of the aerodynamic load is insignificant. In addition, the rocket engine on the RLV has a throttling capability with a limited range of thrust magnitude as follows:
\begin{equation} \label{eq:ptlim}
    \begin{gathered}
         T_{min}\left(1+\mu_T\right) -  P_e A_{exit} \leq \lVert \boldsymbol{T} \rVert \leq T_{max}\left(1-\mu_T\right) - P_e A_{exit}
    \end{gathered}
\end{equation}
In Eq.~\eqref{eq:ptlim}, the parameter $\mu_T$ denotes a thrust margin and serves as a design parameter. For a class of optimal control problems aimed at minimizing fuel consumption under limited thrust, the thrust magnitude profile typically exhibits a bang-bang characteristic, as detailed in~\cite{LuPDG}, where the thrust magnitude instantly moves between its lower and upper bounds. Introducing a tighter thrust range through $\mu_T$ provides additional control authority in the thrust magnitude for the tracking phase to utilize, even when the solution to Problem $\mathcal{P}$ results in either maximum or minimum thrust. Although this approach inherently compromises fuel optimality~\cite{LuPDG2}, it is reported to be essential for achieving stable and robust tracking performance in PDG applications~\cite{LuPDG2, Scharf2017, Ridderhof2021}. Moreover, to avoid a drastic change in the thrust profile, which is not possible with typical rocket engines, the rate of change in thrust magnitude is also constrained as follows:
\begin{equation} \label{eq:ptchange}
    \begin{gathered}   
    - \dot{T}_{lim} \leq \frac{d}{dt}\lVert \boldsymbol{T} \rVert \leq \dot{T}_{lim}
    \end{gathered}
\end{equation}
Lastly, based on Assumption~\ref{asm:thrustalign}, a constraint on the thrust angle is implemented to prevent the vehicle from experiencing a horizontal orientation and to ensure a smooth maneuver in the terminal phase.
\begin{equation} \label{eq:ptang}
    \begin{gathered}
    T_z \leq - \lVert \boldsymbol{T} \rVert \cos \theta_{lim}^p\left(t\right)
    \end{gathered}
\end{equation}
In Eq.~\eqref{eq:ptang}, the limit angle $\theta_{lim}^p$ is described as a function of time, since the time-varying limit angle profile is applied to induce the vertical attitude as the vehicle lands, which is described in a later section.

\subsubsection{Optimal Control Problem}
In Problem $\mathcal{P}$, the terminal mass of the vehicle is the performance index to be minimized to result in a fuel-optimal trajectory. With the cost function defined, we formulate the optimal control problem by incorporating the vehicle dynamics and constraints outlined thus far.
\begin{equation} \label{eq:problemp}
\begin{gathered}
\text{Problem} \, \, \mathcal{P} : \quad \minimize_{\boldsymbol{T},t_f, t_c} \quad J = - \,m\left({t_f}\right) \\
 \text{subject to} \quad { \text{Eqs.}~\eqref{eq:pdyn},~\eqref{eq:pinit},~\eqref{eq:pterm},~\eqref{eq:paload2} \text{ to}~\eqref{eq:ptang} }
\end{gathered}
\end{equation}
It is crucial to note that both the ignition time of the landing burn $t_c$ and the duration of the landing burn $t_f$ are included as optimization variables, which gives the RLV the ability to autonomously determine the engine ignition timing. Hereafter, the optimal solution to Problem $\mathcal{P}$ is denoted by superscript ${}^*$, e.g., $t_f^*$.

\subsection{Tracking Problem}
The objective of the tracking problem is to find the optimal control sequence that maximizes tracking accuracy while satisfying given path constraints. Contrary to Problem $\mathcal{P}$, the OCP for the tracking, referred to as Problem $\mathcal{T}$, has a fixed length prediction horizon $t_h$.

\subsubsection{Vehicle Dynamics}
\begin{figure}[!htb]
    \centering
    \includegraphics[width=5cm]{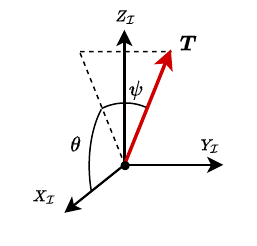}
    \caption{Thrust vector representation.}
    \label{fig:tvec}
\end{figure}

In Problem $\mathcal{P}$, the control variable is the thrust vector generated by the rocket engine. In Problem $\mathcal{T}$, we employ a 5-DOF dynamics model, wherein the pitch, yaw, and engine throttle are augmented to the state variables. This approach is adopted to circumvent the complexities associated with 6-DOF dynamics while still partially preserving the transient characteristics of both rotational and throttle dynamics.
\begin{equation} \label{eq:tvars}
\begin{gathered}
\boldsymbol{X}  =   \left[ \boldsymbol{r},\boldsymbol{v}, m, \theta, \psi, \Gamma \right]^T, \quad \boldsymbol{U} = \left[ \theta_c, \psi_c, {\Gamma}_c \right]^T,\quad \boldsymbol{Z} = \left[ X, U \right]^T 
\end{gathered}
\end{equation}
Here, $\Gamma \triangleq \lVert \boldsymbol{T} \rVert \in \mathbb{R}$ is the thrust magnitude. As in Eq.~\eqref{eq:tvars}, the control variables are the pitch, yaw, and thrust magnitude command. The vehicle dynamics are identical to Problem $\mathcal{P}$, except for the augmented states whose dynamics are defined as follows:
\begin{equation} \label{eq:tdyn}
    \begin{gathered}
    \dot \theta = \frac{1}{\tau_{\theta}} \left(\theta_c - \theta \right),\quad \dot \psi = \frac{1}{\tau_{\theta}} \left(\psi_c - \psi \right),\quad \dot \Gamma = \frac{1}{\tau_{T}} \left(\Gamma_c - \Gamma \right)
    \end{gathered}
\end{equation}
Similar to previous works~\cite{Zarchan, ChenLag} that incorporated aircraft transient dynamics into the optimal control problem to improve the guidance algorithm in demanding scenarios, this work models the attitudes and throttle of the RLV as first-order linear systems. The time constants $\tau_\theta$ and $\tau_T$ are selected to accurately represent the transient response of the attitude control loop. Consequently, the optimized control sequence $\boldsymbol{U}$ is able to account for the relatively slow dynamics of the RLV, thus improving overall guidance performance. The thrust vector is a function of $\theta$, $\psi$, and $\Gamma$ with a $Y-Z$ rotation sequence as depicted in Fig.~\ref{fig:tvec}.
\begin{equation} \label{eq:ttvec}
    \begin{gathered}
\boldsymbol{T} = \Gamma \begin{bmatrix}
\cos \psi \cos \theta \\
\sin \psi \\
-\cos \psi \sin \theta
\end{bmatrix}
    \end{gathered}
\end{equation}

\subsubsection{Constraints}
The initial condition of the tracking problem is given by the 
states of the vehicle at $t_0 \in \left[0, t_f^*\right] $, which is the time when Problem $\mathcal{T}$ is solved. 
\begin{equation} \label{eq:tinit}
    \begin{gathered}
        \boldsymbol{r}\left(t_0\right) =   \boldsymbol{r}\left(t\right), \quad \boldsymbol{v}\left(t_0\right) =   \boldsymbol{v}\left(t\right), \quad m\left(t_0\right)  =   m\left(t\right),\\
        \theta\left(t_0\right) = \theta\left(t\right), \quad \psi\left(t_0\right) = \psi\left(t\right), \quad \Gamma\left(t_0\right) = \Gamma\left(t\right)
    \end{gathered}
\end{equation}
The terminal condition for this problem is not explicitly defined; instead, position and velocity errors relative to the reference trajectory are minimized through the performance index defined in a later section. Analogous to Problem $\mathcal{P}$, path constraints related to vehicle limits and safety considerations are enforced. 
\begin{equation} \label{eq:taload}
    \begin{gathered}
         g_L\left(\boldsymbol{X}, \boldsymbol{U}\right) = {\boldsymbol{T} \cdot \boldsymbol{v}} + \Gamma \lVert \boldsymbol{v}\rVert   \cos \left( \frac{L_{lim}^t}{\bar q} \right) \leq 0
    \end{gathered}
\end{equation}
\begin{equation} \label{eq:ttlim}
    \begin{gathered}
         T_{min}  -  P_e A_{exit} \leq \Gamma \leq T_{max}  -  P_e A_{exit}
    \end{gathered}
\end{equation}
\begin{equation} \label{eq:ttchange}
    \begin{gathered}
    - \dot{T}_{lim} \leq \frac{d}{dt}\Gamma \leq \dot{T}_{lim}
    \end{gathered}
\end{equation}
\begin{equation} \label{eq:ttang}
    \begin{gathered}
    \frac{\pi}{2} - \theta_{lim}^t \leq \theta \leq \frac{\pi}{2} + \theta_{lim}^t,\\
     - \theta_{lim}^t \leq \psi \leq \theta_{lim}^t
    \end{gathered}
\end{equation}
An additional advantage of the thrust representation in Eq.~\eqref{eq:ttvec} is that it allows the thrust magnitude $\Gamma$ to be independently incorporated into the optimization problem. Consequently, the non-convex and second-order cone constraint presented in Eq.~\eqref{eq:ptlim} is transformed into a linear inequality in Eq.~\eqref{eq:ttlim}. This representation of thrust eliminates the need for SOCP in Problem $\mathcal{T}$. Additionally, the thrust margin $\mu_T$ is lifted compared to Eq.~\eqref{eq:ptlim}, so that Problem $\mathcal{T}$ fully utilizes available thrust. The thrust bounds are applied at the command, whereas the attitude bounds are applied at the states. This distinction is due to the fact that the thrust command is physically bounded by the limitation of the rocket engine. On the contrary, the attitude command has no such strict bound except for the attitude error limit from a practical standpoint, which is usually managed inside the attitude control loop.

\subsubsection{Optimal Control Problem}
Before detailing the performance index of Problem $\mathcal{T}$, the state error is defined as follows.
\begin{equation} \label{eq:xerr}
    \begin{aligned}
       \boldsymbol{X}_{e}\left(t\right) &\triangleq \boldsymbol{X}^*\left(t\right) - \boldsymbol{X}\left(t\right), \quad \text{for } t \in \left[ t_0, t_f^t \right] \\ 
       \boldsymbol{X}^*\left(t\right) &= \left[ \boldsymbol{r}^*\left(t\right), \boldsymbol{v}^*\left(t\right),\theta^*\left(t\right),\psi^*\left(t\right),\Gamma^*\left(t\right), m^*\left(t\right)\right]^T
    \end{aligned}
\end{equation}
where $t_f^t \triangleq t_0 + t_h$. In Eq.~\eqref{eq:xerr}, $\boldsymbol{X}^*$ represents the state variables that define the reference trajectories determined by solving the planning problem. It is worth noting that although the attitude angles are not directly optimized in Problem $\mathcal{P}$, they can be readily computed from the optimal thrust profile $\boldsymbol{T}^*$ under Assumption~\ref{asm:thrustalign}. The performance index is given in the form of a linear quadratic tracking law as:
\begin{equation} \label{eq:tj}
    \begin{gathered}
        J = \boldsymbol{X}_e\left(t_f^t\right)^T \boldsymbol{W}_{F} \boldsymbol{X}_e\left(t_f^t\right) + \int_{t_0}^{t_f^t} \boldsymbol{X}_e\left(t\right)^T \boldsymbol{W}_{Q} \boldsymbol{X}_e\left(t\right) + \boldsymbol{\Delta}\left(t\right)^T \boldsymbol{W}_R \boldsymbol{\Delta}\left(t\right) dt
    \end{gathered}
\end{equation}
where $\boldsymbol \Delta \triangleq \left[ \dot{\theta}, \dot{\psi}, \dot{\Gamma} \right]^T$, $\boldsymbol{W}_F, \boldsymbol{W}_Q \in \mathbb{S}^{10}_+$, and $\boldsymbol{W}_F \in \mathbb{S}^{3}_+$. In the performance index, the time derivatives of attitude angles and engine throttle are regulated to control the demands on vehicle maneuverability and engine throttle, thereby adjusting the strain on control systems: TVC actuators and throttle valves. Even though $\boldsymbol{\Delta}$ includes the state derivatives, it can be equivalently expressed as a linear function of states and controls as in Eq.~\eqref{eq:tdyn}. With this, we establish the formulation of the optimal control problem for tracking.
\begin{equation} \label{eq:problemt}
\begin{gathered}
\text{Problem} \, \, \mathcal{T} : \quad \minimize_{\boldsymbol{U}} \quad J \,\, \text{ in Eq.~\eqref{eq:tj}}\\
 \text{subject to} \quad { \text{Eqs.}~\eqref{eq:pdyn},~\eqref{eq:tdyn},~\eqref{eq:tinit} \text{ to}~\eqref{eq:ttang} }
\end{gathered}
\end{equation}
\begin{remark} \label{remark:ttime}
The maximum value of $t_f^t$ is bounded by $t_f^*$. That is, if the prediction horizon $t_f^t$ reaches $t_f^*$, $t_h$ is shrunken as $t_h = t_f^* - t_0$.
\end{remark}

\section{Sequential Convex Programming}\label{sec3}
As outlined in the previous section, both Problem $\mathcal{P}$ and Problem $\mathcal{T}$ feature nonlinear dynamics and nonconvex constraints. To address these OCPs using SCP, convexification and parameterization are necessary, which are the focal points of this section. Initially, the reformulation of Problem $\mathcal{P}$ is presented, which relaxes the problem and allows for the incorporation of flight time into the optimization problem. Subsequently, the formulation of the sub-problems and the SCP algorithm are detailed.

\subsection{Reformulation of Problem \texorpdfstring{$\mathcal{P}$}{P}}
The thrust magnitude constraint in Eq.~\eqref{eq:ptlim} for Problem $\mathcal{P}$ is non-convex due to the existence of a lower bound on the $L_2$ norm. To improve the convergence of SCP by minimizing the approximation errors from linearization, a relaxation is applied to convexify Eq.~\eqref{eq:ptlim}. The thrust magnitude $\Gamma$ is added to the control variables as:
\begin{equation} \label{eq:pu}
\begin{gathered}
\boldsymbol{U} = \left[ \boldsymbol{T}, \Gamma \right]^T,\quad \Gamma \in \mathbb{R}
\end{gathered}
\end{equation}
Then, Equation~\eqref{eq:ptlim} can be relaxed into the intersection of the following two constraints:
\begin{equation} \label{eq:plossless}
\begin{gathered}
\lVert \boldsymbol{T} \rVert \leq \Gamma 
\end{gathered}
\end{equation}
\begin{equation} \label{eq:ptlim2}
\begin{gathered}
T_{min}\left(1+\mu_T\right) -  P_e A_{exit} \leq \Gamma \leq T_{max}\left(1-\mu_T\right) - P_e A_{exit}
\end{gathered}
\end{equation}
This relaxation of the non-convex control constraint into the convex constraint has been widely used in related works~\cite{Acikmese2007, Dueri2017, Jung2023, yang2024}. The proof of exact relaxation, which guarantees the activeness of Eq.~\eqref{eq:plossless} for all time instances, has been conducted for various OCPs~\cite{Acikmese2007, harris2014, Dueri2017, Jung2023, yang2024}. However, this proof does not exist when nonlinear dynamics and multiple nonlinear path constraints exist as in Problem $\mathcal{P}$. Nevertheless, throughout the extensive numerical experiments in this work, the exact relaxation is empirically verified. In terms of the expression of inequality constraints, it is important to note that $\lVert \boldsymbol{T} \rVert$ in Eqs.~\eqref{eq:paload2} to \eqref{eq:ptang} is substituted by $\Gamma$.

Also, Problem $\mathcal{P}$ is a free-final-time problem where the flight time $t_f$ is part of the optimization variable. Deciding the optimal $t_f$ is especially crucial for PDG problems since it majorly determines the fuel consumption. It is not straightforward to include the flight time in the parameter optimization. Therefore, we employ the concept of the time dilation through the introduction of the time variable $\eta$ and subsequent normalized time $\tau = t/\eta \in \left[0,1\right]$. Then, we rewrite the dynamics in Eq.~\eqref{eq:pdyn} to express the time derivatives with respect to $\tau$.
\begin{equation} \label{eq:pdynnorm}
\begin{gathered}
\dot{\boldsymbol{X}} = \boldsymbol{f}^p\left(\boldsymbol{X},\boldsymbol{U}\right) \rightarrow \boldsymbol{X}' = \eta \boldsymbol{f}^p\left(\boldsymbol{X},\boldsymbol{U}\right)
\end{gathered}
\end{equation}
The superscript $'$ denotes the time derivative with respect to $\tau$ and $\boldsymbol{f}^p$ is the dynamics of Problem $\mathcal{P}$ in Eq.~\eqref{eq:pdyn}. Now, by augmenting $\eta$ into the optimization variable, we can directly optimize the flight time $t_f$. The constraint in Eq.~\eqref{eq:ptchange} should also be rewritten since it involves derivatives with respect to time.
\begin{equation} \label{eq:ptchange2}
    \begin{gathered}
    - \dot{T}_{lim} \leq  \eta \Gamma' \leq \dot{T}_{lim}
    \end{gathered}
\end{equation}
Hereafter, the reformulated Problem $\mathcal{P}$ is referred to as Problem $\mathcal{P}^R$, which can be described as follows.
\begin{equation} \label{eq:problempr}
\begin{gathered}
\text{Problem} \, \, \mathcal{P}^R : \quad \minimize_{\boldsymbol{T},\Gamma,\eta, t_c} \quad J = - \,m\left({\tau = \eta }\right) \\
 \text{subject to} \quad { \text{Eqs.}~\eqref{eq:pinit}, \eqref{eq:pterm},~\eqref{eq:paload2},~\eqref{eq:ptang},~\eqref{eq:plossless}\text{ to}~\eqref{eq:ptchange2}}
\end{gathered}
\end{equation}

\subsection{Linearization}
To alleviate the remaining non-convexity in the dynamics and constraints in both problems, we utilize linearization through first-order Taylor expansion to approximate the nonlinear functions with linear affine functions. The linearization of the dynamics varies for each problem due to the existence of the time variable $\eta$. For Problem $\mathcal{P}$, we apply Taylor expansion to Eq.~\eqref{eq:pdynnorm} with respect to both $\eta$ and $\boldsymbol{Z}$.
\begin{equation} \label{eq:pdynlin}
\begin{gathered}
\boldsymbol{X}' \approx \bar \eta \left. \frac{\partial \boldsymbol{f}^p }{\partial \boldsymbol{Z}} \right|_{\bar  {\boldsymbol{Z}}} \left( {\boldsymbol{Z}} - \bar  {\boldsymbol{Z}} \right) + \boldsymbol{f}^p \left(\bar {\boldsymbol{Z}} \right)\eta = \boldsymbol{\bar A}^p \boldsymbol{Z} +  \boldsymbol{\bar C}^p \eta +  \boldsymbol{\bar D}^p
\end{gathered}
\end{equation}
The bar over a variable indicates that the variable is the reference point of the linearization. Conversely, in Problem $\mathcal{T}$, we linearize the dynamics in Eqs.~\eqref{eq:pdyn} and~\eqref{eq:tdyn} only for $\boldsymbol{X}$, since it is a fixed-final-time problem and has a control-affine form with a constant effectiveness matrix, $\boldsymbol{B}^t$.
\begin{equation} \label{eq:tdynlin}
\begin{gathered}
\dot {\boldsymbol{X}} = \boldsymbol{f}^t \left(\boldsymbol{X} \right) + \boldsymbol{B}^t \boldsymbol{U}, \\
\dot {\boldsymbol{X}} \approx \boldsymbol{f}^t \left(\bar{\boldsymbol{X}} \right) +  \left. \frac{\partial \boldsymbol{f}^t }{\partial \boldsymbol{X}} \right|_{\bar  {\boldsymbol{X}}} \left( {\boldsymbol{X}} - \bar  {\boldsymbol{X}} \right) + \boldsymbol{B}^t \boldsymbol{U} = \boldsymbol{\bar A}^t \boldsymbol{X} +  \boldsymbol{B}^t \boldsymbol{U} + \boldsymbol{\bar D}^t
\end{gathered}
\end{equation}
\begin{equation} \label{eq:tB}
\begin{gathered}
\boldsymbol{B}^t = \left[ \begin{array}{c} 
\boldsymbol{0}_{7 \times 3}   \\
diag\left(\frac{1}{\tau_{\theta}},\frac{1}{\tau_{\theta}},\frac{1}{\tau_{T}}\right)
\end{array}\right]
\end{gathered}
\end{equation}
where $\boldsymbol{0}_{7 \times 3}$ denotes an $\mathbb{R}^{7\times 3}$ matrix with zero entries and $diag\left(\cdot\right)$ is the diagonal matrix with zero elements except for the diagonal components given by entries in parentheses. The remaining nonlinearity lies in the initial condition of Problem $\mathcal{P}$ and the aerodynamic load limit for both problems. Equation~\eqref{eq:pinit} can be approximated as follows.
\begin{equation} \label{eq:pinitlin}
    \begin{gathered}
\boldsymbol{r}\left(0\right) \approx \boldsymbol{K}_r \left[-\bar{t}_c^2 , 0 , 1 \right]^T + \boldsymbol{K}_r \left[2\bar{t}_c , 1, 0 \right]^T t_c \\
\boldsymbol{v}\left(0\right) \approx \boldsymbol{K}_v \left[-\bar{t}_c^2 ,0 , 1 \right]^T + \boldsymbol{K}_v \left[2\bar{t}_c , 1, 0 \right]^T t_c
    \end{gathered}
\end{equation}
Lastly, the aerodynamic load constraints in Eqs.~\eqref{eq:paload2} and~\eqref{eq:taload} are linearized as follows.
\begin{equation} \label{eq:aloadlin}
    \begin{gathered}
         g_L\left(\boldsymbol{X}, \boldsymbol{U}\right) \approx  g_L\left(\boldsymbol{\bar X}, \boldsymbol{\bar U}\right) +  \left. \frac{\partial g_L}{\partial \boldsymbol{X}} \right|_{\bar  {\boldsymbol{X}}} \left( {\boldsymbol{X}} - \bar  {\boldsymbol{X}} \right)  +  \left. \frac{\partial g_L}{\partial \boldsymbol{U}} \right|_{\bar  {\boldsymbol{U}}} \left( {\boldsymbol{U}} - \bar  {\boldsymbol{U}} \right) = \boldsymbol{\bar{G}_{X}}\boldsymbol{X} +  \boldsymbol{\bar{G}_{U}}\boldsymbol{U} + \boldsymbol{\bar H} \leq 0
    \end{gathered}
\end{equation}
It is worth pointing out that the constraint in Eq.~\eqref{eq:ptchange2} is also nonlinear since it involves multiplication between optimization variables, but we fix the value of $\eta$ by a reference value $\bar \eta$ to make it a linear constraint by lagging technique~\cite{XinfuLagging}. With the convexification process encompassing the reformulation and linearization discussed thus far, Problem $\mathcal{P}$ and Problem $\mathcal{T}$ can be approximated as convex problems.

\subsection{Parameterization and Soft Trust-Region}
To finalize the formulation of sub-problems for SCP, the main algorithm for solving the OCPs proposed in this work, the convexified problems should be parameterized into a format manageable by numerical solvers. First, the continuous time domain of $\tau$ or $t$ is dicretized into the equally spaced nodes with $N$ intervals.
\begin{equation} \label{eq:timenode}
    \begin{gathered}
       \text{Problem $\mathcal{P^R}$: } \tau \in \left[0, \eta \right] \rightarrow \tau_k = \frac{\bar{\eta}}{N}k \\
       \text{Problem $\mathcal{T}$: } t \in \left[t_0, t_f\right] \rightarrow t_k = t_0 + \Delta t k, \ \Delta t \triangleq \frac{t_f-t_0}{N}k,\\
       \text{for} \ k \in \mathcal{K} = \{ 0,1, \ldots, N\}
    \end{gathered}
\end{equation}
Hereafter, the subscript $k$ denotes that the corresponding variable is related to the $k$-th time node. The number of time nodes $N$ may differ between the planning and tracking problems; however, this distinction is omitted here for the sake of brevity. To transform the continuous ordinary differential equation into equality constraints, an implicit integration method, specifically the trapezoidal rule, is utilized. By the trapezoidal rule, the state dynamics can be discretized into a linear relationship among the optimization variables at two neighboring time nodes.
\begin{equation} \label{eq:ptrapdyn}
    \begin{gathered}
       \text{Problem $\mathcal{P^R}$: } X_{k+1} - X_{k} = \frac{1}{2N} \left(\bar{\boldsymbol A}^p_k \boldsymbol{Z}_k +  \bar{\boldsymbol C}^p_k \eta +  \boldsymbol{\bar D}^p_k + \bar{\boldsymbol A}^p_{k+1} \boldsymbol{Z}_{k+1} +  \bar{\boldsymbol C}^p_{k+1} \eta +  \bar{\boldsymbol D}^p_{k+1} \right), \\ \text{for} \ k \in \mathcal{K}^- = \{ 0,1, \ldots, N-1\}
    \end{gathered}
\end{equation}
\begin{equation} \label{eq:ttrapdyn}
    \begin{gathered}
       \text{Problem $\mathcal{T}$: } X_{k+1} - X_{k} = \frac{\Delta t}{2}  \left( \bar{\boldsymbol A}^t_k \boldsymbol{X}_k +  \boldsymbol{B}^t_k \boldsymbol{U}_k + \bar{\boldsymbol D}^t_k + \bar{\boldsymbol A}^t_{k+1} \boldsymbol{X}_{k+1} +  \boldsymbol{B}^t_{k+1} \boldsymbol{U}_{k+1} + \bar{\boldsymbol D}^t_{k+1}\right) \\ \text{for} \ k \in \mathcal{K}^-
    \end{gathered}
\end{equation}
The thrust rate constraint in Eqs.~\eqref{eq:ttchange} and~\eqref{eq:ptchange2} also involves the time derivative, which can be parameterized as below.
\begin{equation} \label{eq:ptchangedisc}
    \begin{gathered}
      \text{Problem $\mathcal{P^R}$: } -\dot{T}_{lim} \leq \frac{\bar{\eta}}{N} \lvert \Gamma_{k+1} -\Gamma_k \rvert \leq \dot{T}_{lim},\quad \text{for} \ k \in \mathcal{K}^-
    \end{gathered}
\end{equation}
\begin{equation} \label{eq:ttchangedisc}
    \begin{gathered}
      \text{Problem $\mathcal{T}$: } -\dot{T}_{lim} \leq \lvert \Gamma_{k+1} -\Gamma_k \rvert \Delta t \leq \dot{T}_{lim},\quad \text{for} \ k \in \mathcal{K}^-
    \end{gathered}
\end{equation}
Lastly, by the trapezoidal rule, the cost function of Problem $\mathcal{T}$ can be parameterized as the sum of quadratic functions in $\boldsymbol{Z}_k$.
\begin{equation} \label{eq:ttrapcost}
    \begin{gathered}
       J^t = \boldsymbol{X}_{e,N}^T \boldsymbol{W}_F \boldsymbol{X}_{e,N} + \frac{t_h}{N}\sum_{k=0}^{N} w_k \left( \boldsymbol{X}_{e,k}^T \boldsymbol{W}_{Q} \boldsymbol{X}_{e,k} + \boldsymbol{\Delta}_{k}^T\boldsymbol{W}_R \boldsymbol{\Delta}_{k} \right)
    \end{gathered}
\end{equation}
Here, the parameter $w_k$ is the coefficient used to express the integral over time using the trapezoidal rule.
\begin{equation} \label{eq:trapw}
    \begin{gathered}
    w_k     =   
        \begin{cases}
            \text{if } k \in \{ 0 , N \}:  & {1}/{2} \\
            \text{else } : & 1
        \end{cases}
    \end{gathered}
\end{equation}

Due to the approximations made through linearization in the convex sub-problem, as mentioned earlier, a trust-region is required to maintain stable and well-defined parameter optimization during SCP. To address the challenge of selecting an appropriate fixed size for the hard trust-region of each optimization variable, we adopt a soft trust-region approach, previously utilized in works~\cite{SzmukAIAA, Szmuk6DSTC}. This approach involves augmenting the cost function with the weighted quadratic difference between the reference point and the optimization variables.
\begin{equation} \label{eq:softtrust}
    \begin{gathered}
        J_{tr}  =   \sum_{k=0}^{N} \left(\Delta \boldsymbol{Z}_k\right)^T \boldsymbol{W}_{tr}\ \Delta \boldsymbol{Z}_k
    \end{gathered}
\end{equation}
where $\Delta \boldsymbol{Z}_k \triangleq \boldsymbol{Z}_k - \bar{\boldsymbol{Z}}_k$. The matrix $\boldsymbol{W}_{tr}$ is a symmetric positive semi-definite matrix with dimensions identical to $\boldsymbol{Z}$ and is a part of the design variables. With this, we have completed the formulation of the sub-problems for Problem $\mathcal{P}$ and Problem $\mathcal{T}$, which are named as Problem $\mathcal{P}^R_2$ and Problem $\mathcal{T}_2$, respectively.
\begin{equation} \label{eq:problempr2}
\begin{gathered}
\text{Problem} \, \, \mathcal{P}^R_2 : \quad \minimize_{t_c, \eta, \boldsymbol{Z}_{k\in \mathcal{K}}} \quad J = - \,m_N + J_{tr} \\
 \text{subject to} \quad \begin{matrix}
     \text{Eqs. } \eqref{eq:pterm},\eqref{eq:pinitlin} & \text{Boundary Condition} \\ \text{Eqs. } \eqref{eq:ptrapdyn}, \eqref{eq:ptchangedisc} & \text{for } k \in \mathcal{K}^-\\
     \text{Eqs. } \eqref{eq:ptang}, \eqref{eq:plossless}, \eqref{eq:ptlim2}, \eqref{eq:aloadlin} & \text{for } k \in \mathcal{K}
 \end{matrix}
\end{gathered}
\end{equation}
\begin{equation} \label{eq:problemt2}
\begin{gathered}
\text{Problem} \, \, \mathcal{T}_2 : \quad \minimize_{\boldsymbol{Z}_{k\in \mathcal{K}}} \quad J = J^t + J_{tr}, \quad J^t \text{ in Eq.~\eqref{eq:ttrapcost} }\\
 \text{subject to} \quad \begin{matrix}
     \text{Eqs.}~\eqref{eq:tinit} & \text{Boundary Condition} \\ \text{Eqs.}~\eqref{eq:ttlim},~\eqref{eq:ttrapdyn},~\eqref{eq:ttchangedisc} & \text{for } k \in \mathcal{K}^-\\
     \text{Eqs.}~\eqref{eq:ttang},~\eqref{eq:aloadlin} & \text{for } k \in \{ 1,2,\ldots, N \}
 \end{matrix}
\end{gathered}
\end{equation}
The resulting sub-problem $\mathcal{P}^R_2$ is an SOCP problem due to the the existence of the SOC constraint in Eq.~\eqref{eq:plossless}. In contrast, the sub-problem $\mathcal{T}_2$ is a QP problem. It is crucial to note that for Problem $\mathcal{T}_2$, attitude limits and aerodynamic load constraints are not imposed at the first node, corresponding to the initial time $t_0$. Unlike Problem $\mathcal{P}^R_2$, the thrust and attitudes are state variables in Problem $\mathcal{T}_2$, subject to initial conditions determined by the current vehicle states. Therefore, we omit the aforementioned path constraints on the first node.

\subsection{Algorithm}
\begin{algorithm}[tb]
    \caption{ Sequential Convex Programming}
    \label{alg:SCP}
    \begin{tabular}{ll}
    1. \textbf{Initialize}& \textbf{Obtain} initial reference variables: $\left[ \ \{ \bar{t}_c, \bar{\eta}, \bar{\boldsymbol{Z}}_{k\in \mathcal{K}}\} \ || \ \bar{\boldsymbol{Z}}_{k\in \mathcal{K}}\ \right]$   \\
 & \textbf{Set} $i=1$, where $i$ is the SCP iteration number \\
2. \textbf{Solve Sub-problem} & Compute linearization matrices   \\
 & Optimize sub-problem  $\left[ \ \mathcal{P}^R_2 \ || \ \mathcal{T}_2 \ \right] $ by $\left[ \ \text{SOCP}\ ||\ \text{QP }\ \right] $ solver.  \\ 
 & Retrieve optimized solution:  $\left[ \ \{ {t}_c^i, {\eta}^i, {\boldsymbol{Z}}_{k\in \mathcal{K}}^i\} \ || \ {\boldsymbol{Z}}_{k\in \mathcal{K}}^i\ \right]$  \\
 3. \textbf{Check Convergence} & Evaluate $J_{tr} < \epsilon_{SCP}$  \\ 
 & $\quad$ \textbf{If} satisfied, $\left[ \ \{ \bar{t}_c^i, {\eta}^i, {\boldsymbol{Z}}_{k\in \mathcal{K}}\}^i \rightarrow \{{t}_c^*, {\eta}^*, {\boldsymbol{Z}}_{k\in \mathcal{K}}^*\} \ || \ {\boldsymbol{Z}}_{k\in \mathcal{K}}^i\rightarrow {\boldsymbol{Z}}_{k\in \mathcal{K}}^* \ \right]$ and \textbf{Terminate}  \\
 & $\quad$ \textbf{Else}, update the reference point  $\left[ \ \{ \bar{t}_c^i, {\eta}^i, {\boldsymbol{Z}}_{k\in \mathcal{K}}\}^i \rightarrow \{\bar{t}_c, \bar{\eta}, \bar{\boldsymbol{Z}}_{k\in \mathcal{K}}\} \ || \ {\boldsymbol{Z}}_{k\in \mathcal{K}}^i\rightarrow \bar{\boldsymbol{Z}}_{k\in \mathcal{K}} \ \right]$,\\
 &\quad \quad \textbf{Set}  $i=i+1$ and \textbf{Do} from 2. 
\end{tabular}
\end{algorithm}
The SCP algorithm utilized in this work is detailed in Algorithm~\ref{alg:SCP}. The superscript $i$ denotes that the corresponding variable is part of the solution at the $i$-th SCP iteration. The SCP algorithm requires an initial guess for the reference trajectory to start the optimization loop. For Problem $\mathcal{P}^R_2$, once the appropriate guesses of ignition time $t_c$ and flight duration $\eta$ are given, the initial guesses for position and velocity are computed as simple linear interpolations between initial and final points. 
\begin{equation} \label{eq:piguessx}
    \begin{gathered}
        \bar{\boldsymbol r}_k   =  \boldsymbol{r}_0 + \left(\boldsymbol{r}_f - \boldsymbol{r}_0 \right)\tau_k, \quad \bar{\boldsymbol v}_k   =  \boldsymbol{v}_0 + \left(\boldsymbol{v}_f - \boldsymbol{v}_0 \right)\tau_k, \quad \text{for } k \in \mathcal{K}
    \end{gathered}
\end{equation}
For the majority of the RLV missions, the primary velocity component that must be nullified is aligned with the vertical axis. Consequently, the initial guesses for the control inputs are determined as follows:
\begin{equation} \label{eq:piguessu}
    \begin{gathered}
        \bar{\boldsymbol{U}}_k     =  \left[ 0, 0, -\frac{T_{min}+T_{max}}{2} - A_e P_{atm,k}, \frac{T_{min}+T_{max}}{2} - A_e P_{atm,k} \right]^T, \quad \text{for } k \in \mathcal{K}
    \end{gathered}
\end{equation}
The corresponding mass profile can simply be obtained by integrating $\Gamma_k$ from Eq. \eqref{eq:piguessu}. The initial guess for Problem $\mathcal{T}_2$ can easily be derived from the reference trajectory or the previous optimization result since it is recursively solved throughout the descent. 
\begin{remark}
The optimization variables $\boldsymbol{Z}$ of each SCP sub-problem are normalized based on the minimum and maximum values of each state and control variable to enhance the numerical stability.
\end{remark}

\section{Model Predictive Guidance}\label{sec4}
This section elaborates on the core concept of the proposed model predictive guidance in detail. Initially, an overview of the overall architecture of the proposed algorithm is presented. Subsequently, the specifics regarding the selection of design variables are detailed. Finally, the feasibility characteristics of Problem $\mathcal{T}$ are examined.

\subsection{Proposed Framework}
To achieve near closed-loop guidance performance through trajectory optimization, rapid solution updates are necessary to ensure that the optimized trajectory reflects the vehicle’s current states. Computing the fuel-optimal trajectory requires incorporating the entire path from the current state to the landing site, thus enlarging the problem size. This issue can be worsened as the model fidelity of the planning problem expands from 3-DOF to 6-DOF, as shown in~\cite{Szmuk6DSTC, Sagliano2023a}. Even though the computational load can decrease as the vehicle nears the landing site, the inherent computational intensity may become an obstacle to achieving a fast update rate of the solution throughout the flight. Moreover, the trajectory planning itself tends to become numerically unstable in the terminal phase of the guidance~\cite{Lu2023} for the class of fuel-optimal problems. Since the maximum thrust magnitude is reached at the end of such problems, there is no additional control authority to compensate for trajectory deviations, which can render the problem infeasible.
\begin{figure}[!htb]
    \centering
    \includegraphics[width=6cm]{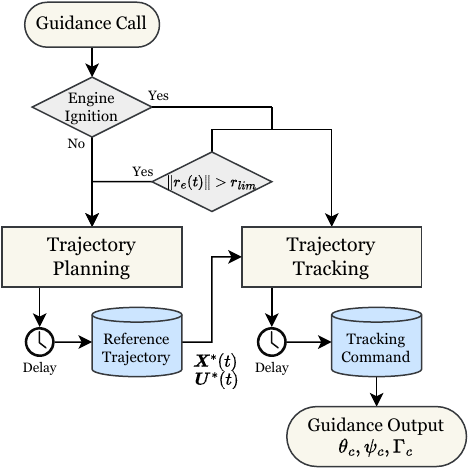}
    \caption{Flowchart of the proposed guidance algorithm.}
    \label{fig:flowchart}
\end{figure}

Landing guidance through trajectory optimization is typically achieved by either recalculating the fuel-optimal trajectory and feed-forwarding the optimal thrust profile~\cite{WangMPC, Guadagnini2022}, or by determining an optimal trajectory initially and subsequently tracking it~\cite{bollino2006, Scharf2017, Sagliano2023a}. This work introduces a hybrid approach that seeks a practical balance in the numerical strategy. The guidance system consists of two main elements: trajectory planning, achieved by solving Problem $\mathcal{P}^R$, and determining the optimal tracking command by solving Problem $\mathcal{T}$. By employing a trajectory optimization-based tracking law, the proposed algorithm relies less on trajectory planning for optimal guidance command determination. Additionally, Problem $\mathcal{P}^R$ employs 3-DOF dynamics to lighten the computational load, while delegating the consideration of model fidelity to Problem $\mathcal{T}$. The flowchart of the proposed algorithm is illustrated in Fig.~\ref{fig:flowchart}. The latest solution from trajectory planning serves as the reference for the tracking algorithm, and the optimized pitch, yaw, and throttle commands ($\theta_c, \psi_c, \Gamma_c$) are the final output of the proposed guidance.

In this work, we plan the fuel-optimal trajectory for only two occasions: initial planning during the aerodynamic descent prior to the landing burn, and instances where the positional tracking error exceeds the prescribed threshold $r_{lim}$. The trajectory planning through Problem $\mathcal{P}^R$ does not require an immediate solution update. Instead, an appropriate initiation altitude for the trajectory planning can be chosen to safely obtain the optimal reference trajectory prior to the expected ignition time $t_c$. Moreover, updating the reference trajectory is not strictly necessary when the trajectory deviation is not significantly large, since a newly optimized trajectory will be similar to the previous trajectory based on the principle of optimality~\cite{liberzon2012}. Therefore, we re-plan the reference trajectory when the position tracking error exceeds the prescribed limit $r_{lim}$. For re-planning during the landing burn, the initial condition of Problem $\mathcal{P}^R$ is substituted with the current state of the vehicle without involving $t_c$ in Eq.~\eqref{eq:pinit}. 
\begin{equation} \label{eq:pinit2}
    \begin{gathered}
\boldsymbol{r}\left(0\right) =  \boldsymbol{r}\left(t\right), \quad \boldsymbol{v}\left(0\right) =  \boldsymbol{v}\left(t\right),\quad m\left(0\right)  =   m\left(t\right)
    \end{gathered}
\end{equation}

For the task of trajectory tracking, we solve Problem $\mathcal{T}$ using the SCP algorithm to obtain updated optimal control sequences for attitudes and engine throttle. These sequences are aimed at minimizing the tracking error while satisfying critical path constraints. Subsequently, the guidance commands are sent to the control loop by time interpolating the latest discretized solution of Problem $\mathcal{T}$. To enhance the guidance performance, we devise a combination of lift force compensation and augmented states for attitudes and throttle. This approach increases the model fidelity of Problem $\mathcal{T}$. Despite the increased number of states compared to Problem $\mathcal{P}^R$, the computational load for Problem $\mathcal{T}$ can be lower due to sub-problem $\mathcal{T}_2$ being a QP problem, which is less computationally complex than SOCP~\cite{alizadeh2003}. Furthermore, by adjusting the horizon length $t_h$, the problem size can be tuned to enable rapid solution updates and improved tracking performance. As a result, we recursively solve the tracking problem at a predetermined frequency $\mathcal{F}^t$ throughout the landing burn of the RLV, providing consistent and rapid updates of optimal tracking commands.

\subsection{Lift Force Compensation} 
\begin{figure}[!htb]
    \centering
    \includegraphics[width=9cm]{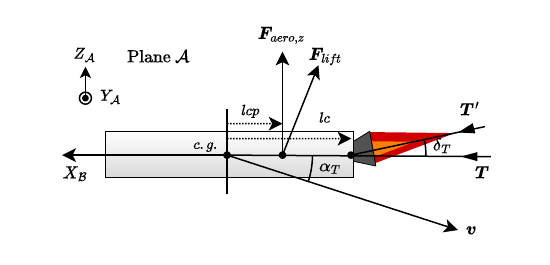}
    \caption{Lift compensation.}
    \label{fig:liftcomp}
\end{figure}

In a real-world scenario, the thrust deflection by TVC governed by the attitude control algorithm exists. As shown in Fig.~\ref{fig:liftcomp}, there is a clear difference between the thrust vector $\boldsymbol{T}$ based on Assumption~\ref{asm:thrustalign} in this work and the actual thrust vector $\boldsymbol{T}'$ with TVC deflection. This difference can result in a significant overestimation or underestimation of the tangential aerodynamic force generated by the angle of attack, which degrades the guidance performance. To consider this perturbation of thrust force without complicating the problem formulation, the lift coefficient is passively modified as follows.
Figure~\ref{fig:liftcomp} depicts the forces exerted on the vehicle from a perspective fixed orthogonally to plane $\mathcal{A}$. First, we model the TVC deflection angle command from the attitude control algorithm in Fig.~\ref{fig:liftcomp} as two parts: 
\begin{equation} \label{eq:tvcaero1}
    \begin{gathered}
    \delta_{T,c}     =   \delta_{T,m} + \delta_{T,A}
    \end{gathered}
\end{equation}
where the variable $\delta_{T,m}$ is the TVC deflection required for the rigid-body maneuver and the variable $\delta_{T,A}$ is for nullifying the aerodynamic moments. These can be computed as:
\begin{equation} \label{eq:tvcaero2}
    \begin{gathered}
    \delta_{T,A}     =   \sin^{-1} \left( {- \frac{M_{aero,y}}{l_c \lVert \boldsymbol{T} \rVert } } \right)
    \end{gathered}
\end{equation}
where the variable $M_{aero,y}$ represents the aerodynamic moment caused by the aerodynamic force $F_{aero,z}$ exerted at the center of pressure about the lateral axis $Y_\mathcal{A}$, as defined in Fig.~\ref{fig:liftcomp}. The origin of this axis is located at the center of gravity of the vehicle.
\begin{equation} \label{eq:tvcaero3}
    \begin{gathered}
    M_{aero,y} = l_{cp} F_{aero,z} = \bar q s_{ref} l_{cp} C_z
    \end{gathered}
\end{equation}
where $C_z \triangleq C_L \cos \alpha_T + C_D \sin \alpha_T$. From Eqs.~\eqref{eq:tvcaero2} and~\eqref{eq:tvcaero3}, the compensated lift force, $\boldsymbol{F}_{lift}'$, that augments the component of $\boldsymbol{T}'$ perpendicular to the velocity vector by $\delta_{T,A}$, can be written independently from the thrust magnitude as follows.
\begin{equation} \label{eq:tvcaero4}
    \begin{gathered}
        \boldsymbol{F}_{lift}' = \left(\bar q C_L s_{ref} + \lVert \boldsymbol{T} \rVert \sin {\left(\delta_{T,A} + \alpha_T\right)} - \lVert \boldsymbol{T} \rVert \sin { \alpha_T } \right)  \hat {\boldsymbol{L}} \\ \approx  \left(\bar q C_L s_{ref} + \lVert \boldsymbol{T} \rVert \sin {\delta_{T,A} } \right)  \hat {\boldsymbol{L}} =  \bar q s_{ref} 
    \left( C_L - \frac{l_{cp}}{l_c}C_z\right)\hat {\boldsymbol{L}} = \bar q C_L' \hat {\boldsymbol{L}}
    \end{gathered}
\end{equation}
Hereafter, the compensated lift coefficient $C_L'$ substitutes $C_L$ in Eqs.~\eqref{eq:pdyn} and~\eqref{eq:tdyn} for computing the lift force. Note that only the value of $l_c$, which is the moment arm length of the TVC, is required to compute $C_L'$, aside from the aerodynamics data of the vehicle. This means that the numerical complexity of solving the OCP is not increased.
\begin{remark} \label{remark:liftcomp}
For a general attitude control law with angular feedback and integral gain, the TVC deflection command can be modeled as in Eq.~\eqref{eq:tvcaero1}, leveraging the disturbance (moment by aerodynamics in this case) rejecting property of the control algorithm.
\end{remark}
\begin{remark}
$\delta_{T,A}$ is usually more dominant in generating lateral acceleration than $\delta_{T,m}$, since $\delta_{T,m}$ accounts for short-period dynamics, whereas $\delta_{T,A}$ deals with the moment bias induced by angle of attack, which persists throughout the flight. Therefore, only $\delta_{T,A}$ is included in the lift compensation.
\end{remark}

\subsection{Design Variables}
For the RLV's safe landing, minimizing terminal position and velocity errors is essential, but ensuring upright terminal attitudes and minimal attitude rates is also crucial. To guide the vehicle towards achieving safe terminal attitudes, we design the thrust tilt angle $\theta_{lim}^p$ for the planning problem without adding extra constraints. For two design variables,  $t_{\theta}$ and the maximum limit of the tilt angle $\bar\theta_{lim}$, we impose time-varying $\theta_{lim}^p$ so that the resulting trajectory has a constant angular acceleration and zero angular rate in the tilt angle at the terminal phase.
\begin{equation} \label{eq:angacc}
    \begin{gathered}
    \theta_{lim}^p\left(t\right) =   
        \begin{cases}
            \text{if } t \leq t_f - t_{\theta}:  & \bar \theta_{lim} \\
            \text{else } : & \frac{1}{2} K_\theta \left( t_f - t \right)^2
        \end{cases}
        , \text{ where} \quad K_\theta = \frac{2\bar \theta_{lim}}{t_\theta^2}
    \end{gathered}
\end{equation}
As the tilt angle approaches vertical at the flight's end by Eq.~\eqref{eq:angacc}, tracking the reference trajectory may result in a vertical attitude for the vehicle upon landing. A longer $t_\theta$ reduces the angular acceleration $K_\theta$ and results in trajectory tracking with less maneuver. However, since extending $t_\theta$ narrows the feasible trajectory space, it should be decided with care.

To robustly track the reference trajectory in the tracking problem, we design the parameters for the path constraints to provide a wider feasible region of variables for Problem $\mathcal{T}$ than Problem $\mathcal{P}^R$. We impose the following condition for a constant $\theta_{lim}^t$.
\begin{equation} \label{eq:ttang2}
    \begin{gathered}
        \bar \theta_{lim} < \theta_{lim}^t
    \end{gathered}
\end{equation}
The aerodynamic load constraint also limits the attitudes of the launch vehicle around the velocity vector. Therefore, $L_{lim}$ for each problem should have the following relationship:
\begin{equation} \label{eq:aloadval}
    \begin{gathered}
        L_{lim}^p <  L_{lim}^t
    \end{gathered}
\end{equation}
For the thrust magnitude, a thrust margin $\mu_T$ with a constant and positive value is imposed on Problem $\mathcal{P}^R$ as in Eq.~\eqref{eq:ptlim}. 

The weight matrices for the cost function of Problem $\mathcal{T}$: $\boldsymbol{W}_{F}$,$ \boldsymbol{W}_{Q}$, and $ \boldsymbol{W}_{U}$, majorly determine the performance and characteristics of the tracking guidance and can reflect the priorities of states by tuning the ratio between diagonal components. The overall ratio between $\boldsymbol{W}_{Q}$ and $\boldsymbol{W}_{U}$ decides the desired tracking agility, as $\boldsymbol{W}_{U}$ penalizes the attitudes and throttle rates. In this work, we aim to match the time of landing as the planned value ($t_f^*$) to preserve fuel optimality. To promote this behavior, we can heavily weight the altitude component of $\boldsymbol{W}_{Q}$, which also helps in retaining the vertical velocity at landing as intended. The values in the terminal cost weight $\boldsymbol{W}_F$ can be utilized as an additional knob to tune the guidance performance. When the prediction horizon does not reach the landing site, $\boldsymbol{W}_F$ can be designed to improve the overall tracking performance. In instances where the prediction horizon reaches the landing site, $\boldsymbol{W}_F$ directly decides the priority between the desired final states. For example, if the RLV mission allows for some horizontal landing position error but requires a soft and upright landing for safety, the velocity and attitudes component of $\boldsymbol{W}_F$ can be heavily weighted compared to the horizontal position. Lastly, $\boldsymbol{W}_{R}$ can be designed depending on the available strains on each control loop for attitudes or throttle.

\subsection{Feasibility of Problem \texorpdfstring{$\mathcal{T}$}{T}}
The planning problem $\mathcal{P}^R$ can become infeasible during the flight as the trajectory deviation builds up due to various uncertainties. If the feasibility of the tracking problem $\mathcal{T}$ can be ascertained, the proposed guidance provides stable performance regardless of the current feasibility of the planning task. To discuss the feasibility of Problem $\mathcal{T}$, we treat it as a discretized problem, whose solution can be obtained from the converged solution of Problem $\mathcal{T}_2$ through SCP.

Firstly, disregarding the path constraints, the states and controls that satisfy the dynamics constraints can always be found since the dynamics is an autonomous differential equation. Then, we assume that the thrust control has minimal uncertainties, meaning the thrust reported by the dedicated thrust control scheme always falls within the minimum and maximum throttle bounds, as given by engine specifications. This assumption does not eliminate the possibility of the actual thrust being different from the thrust known to the control algorithm. With this assumption, the magnitude change constraint in Eq.~\eqref{eq:ttchangedisc} can be satisfied by $\Gamma_{c,k}, k \in \mathcal{K}$ constrained by Eq.~\eqref{eq:ttlim}.

Then, the remaining constraints on states in Problem $\mathcal{T}$ are the aerodynamic load limit in Eq.~\eqref{eq:taload} and the bounds for each attitude in Eq.~\eqref{eq:ttang}. The cases where infeasibility can occur can be summarized into two cases. First, even though Problem $\mathcal{T}$ provides optimal commands that adhere to the path constraints during the prediction horizon, disturbances and uncertainties in attitude dynamics can cause a violation of the constraints. For this reason, we omit path constraints on $\theta$ and $\psi$ at the first node as described in Eq.~\eqref{eq:problemt2}. Then, the attitude bounds can be satisfied from $k=1$ by unconstrained $\theta_c$ and $\psi_c$ under first-order linear dynamics. It should be noted that drastic guidance commands $\theta_{c}$ and $\psi_c$ at the first and second nodes might occur with a large violation of bounds at the initial condition. In such cases, an appropriate command limit at the input of the attitude control algorithm has to be imposed, which is a common practice in attitude control of launch vehicles.
\begin{figure}[!htb]
    \centering
    \includegraphics[width=7cm]{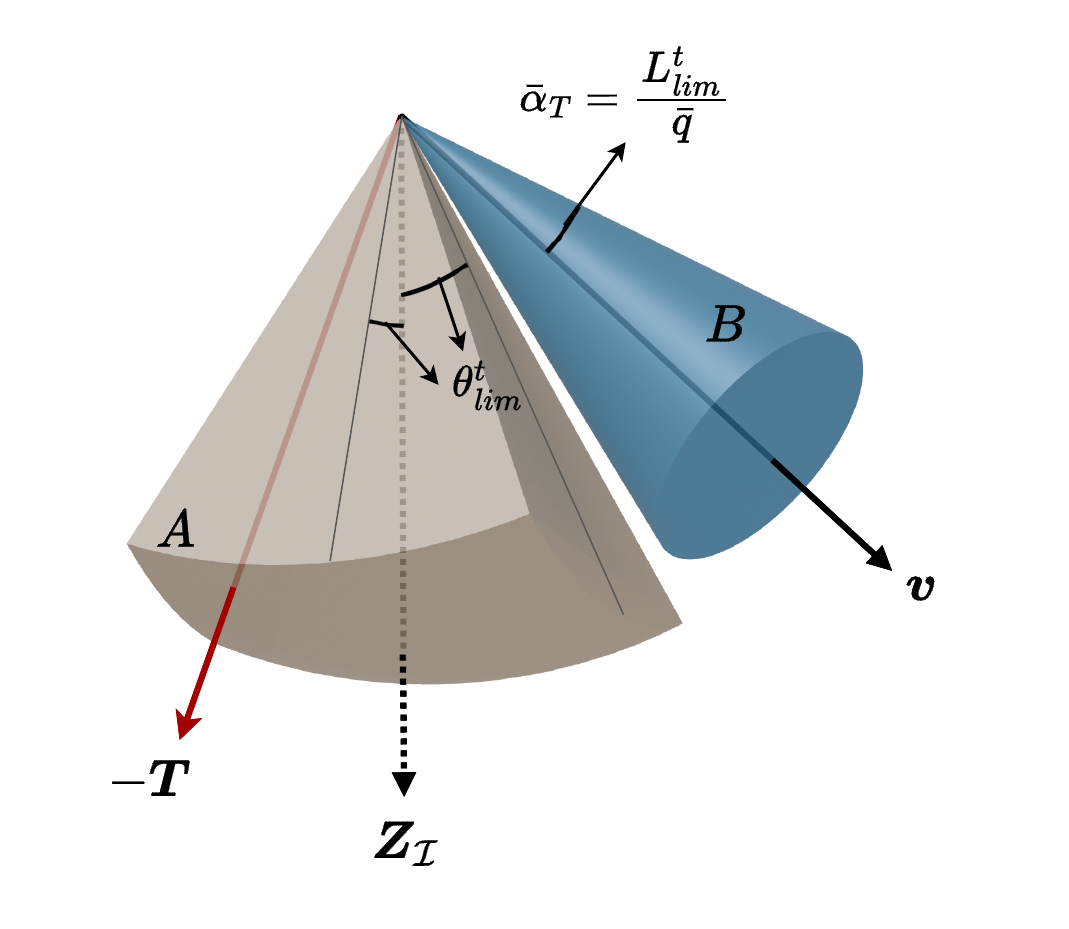}
    \caption{Schematics of feasible regions of $-\boldsymbol{T}$.}
    \label{fig:intersect}
\end{figure}

Second, if the current velocity $\boldsymbol{v}$ is tilted far from the vertical axis, there may be no intersection of feasible regions for $\boldsymbol{T}$ between the aerodynamic load ($B$) and the thrust tilt angle ($A$) as shown in Fig.~\ref{fig:intersect}. To prevent such occasions, the tilt angle bound $\theta_{lim}^t$ should be sufficiently large, considering the possible range of $\boldsymbol{v}$, especially during the instances with high dynamic pressure to ensure that there always exists a non-empty intersection between $A$ and $B$. 

With the preliminary discussions up to this point, Problem $\mathcal{T}$ remains consistently feasible throughout the landing burn and provides well-posed guidance performance when utilizing trajectory optimization. Also, during the extensive numerical experiments conducted in this work, a guidance failure due to the infeasibility of the tracking problem is not observed, which gives empirical confidence in its stability.

\section{Numerical Validation}\label{sec5}
In this section, the numerical results from the 6-DOF simulation program for the proposed guidance algorithm are presented and discussed.

\subsection{Simulation Configuration and Parameters}
We use an RLV vehicle model similar to the first-stage booster of the Falcon 9, equipped with a reaction control system (RCS), aerodynamic fins, and a TVC-controlled rocket engine as control means. In this work, only one rocket engine is assumed to be used for landing. Prior to the landing burn, the RCS and aerodynamic fins control the roll, pitch, and yaw channel. During the landing burn, the control of pitch and yaw angles is handed off to the TVC. Here, a nonlinear 3-loop control law similar to previous work~\cite{Lee2016} is adopted to design a control algorithm for the pitch and yaw channels, eliminating the need for control gain scheduling for time-varying vehicle properties and engine throttle. It is important to note that the designed control law does not explicitly utilize aerodynamic models to compensate for aerodynamic moments. Instead, it transiently nullifies them with the integral gain, which aligns with Remark~\ref{remark:liftcomp}. The control gains are tuned to provide stable control loops considering the performance of the TVC actuator. The time constant of the outer control loop (i.e., attitude control loop), $\tau_\theta$, is approximately $1.0\;{\rm sec}$ and is used in Problem $\mathcal{T}_2$.
\begin{table}[htb]
\centering
\caption{Reusable launch vehicle parameters.}
\label{tbl:LVparam}
\begin{tabular}{cccccc} \hline\hline
Parameter & Value  & Unit    & Parameter       & Value      & Unit      \\ \hline
$l_{total}$ & 42.6   & ${\rm m}$   &  $T_{max} $ &  816          & ${\rm kN}$  \\
$d_{ref}$   & 3.66   & ${\rm m}$   &  $T_{min} $ &  419          & ${\rm kN}$  \\
$s_{ref}$   & 10.52  & ${\rm m^2}$ &$\omega_{TVC}$      &6        & ${\rm Hz}$    \\
$A_e $     & 0.6648  & ${\rm m^2}$ &$\zeta_{TVC}$       &0.707    & $*    $     \\
$m_0 $     & 36.079  & ${\rm ton}$ &$\tau_T$           &0.5       & ${\rm sec}$      \\
$I_{sp} $   & 282    & ${\rm sec}$ &$\dot{T}_{lim}$     &194.7    & ${\rm kN/s}$   \\
\hline\hline
\end{tabular}
\end{table}
\begin{figure}[tb]
    \centering
    \includegraphics[width=6cm]{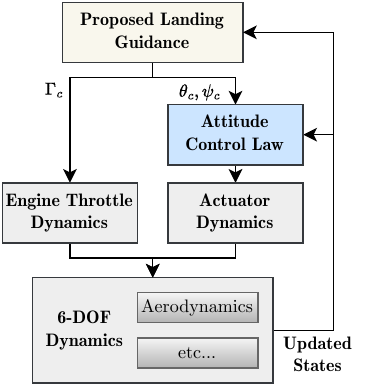}
    \caption{Simulation configuration.}
    \label{fig:sim}
\end{figure}

The 6-DOF simulation program used for numerical experiments in this work is constructed in the MATLAB environment with the overall structure shown in Fig.~\ref{fig:sim}. The TVC actuator is modeled as a second-order linear system characterized by the natural frequency $\omega_{TVC}$ and damping ratio $\zeta_{TVC}$. The engine throttle is modeled as a first-order linear system with a time constant $\tau_T$ and thrust change limit $\dot{T}_{lim}$. The values for the modeling parameters of each component are shown in Table~\ref{tbl:LVparam}. Publicly available aerodynamics data of a first-stage booster in related research~\cite{Simplicio2020} are acquired and adjusted to match the vehicle's size in the development of the aerodynamics module.

To numerically solve sub-problems of $\mathcal{P}^R$ and $\mathcal{T}$ with SCP, we use ECOS~\cite{Domahidi2013} and CQSol, an in-house primal-interior QP solver. Both solvers are written in C language and can be used in MATLAB with compiled performance. CQSol employs an algorithm similar to that of CVXGEN~\cite{Mattingley2012}, with the primary difference being its non-customized nature, allowing it to accept general QP problems. It is worth noting that while ECOS can solve QP problems, it requires additional SOC constraints for the transformation of QP to SOCP. Thus, to fully utilize the simplicity of QP, a dedicated QP solver is employed in this work. The 6-DOF simulation and numerical solvers are executed on a desktop with an intel i7 processor. Lastly, the computational delays for solving each OCP are reflected in the simulation by delaying the update of the solution according to the required solving time.

The parameters for both OCPs and the SCP algorithm can be seen in Table~\ref{tbl:probparam}. The weight matrices that define the cost function of Problem $\mathcal{T}$ are described in Table~\ref{tbl:trackparam}. These matrices are empirically designed to prioritize fuel-optimal performance and soft landing, rather than focusing strictly on precise landing, considering that the landing pads of RLVs have a size of several dozen meters. 

Lastly, Table~\ref{tbl:simIC} shows two initial conditions for the nonlinear simulation. IC 1 represents the ideal case that results in minimal maneuvering and angle of attack for a fuel-optimal trajectory, while IC 2 involves a considerable amount of both maneuvering and $\alpha_T$.
\begin{table}
\centering 
\caption{Problem/Algorithm parameters.}
\label{tbl:probparam}
\begin{tabular}{cccccc} 
\hline\hline
Parameter & Value & Unit & Parameter & Value & Unit \\ 
\hline
   $L_{lim}^p$      & $3.0\times 10^3$  &  ${\rm Pa\cdot rad}$    &   $N^p$       &100 &  *    \\
   $L_{lim}^t$      & $3.5\times 10^3$  &  ${\rm Pa\cdot rad}$   &   $N^t$       &32  &  *    \\
   $t_{\theta}$     & $5$&  ${\rm sec}$       &$t_h$              & $8$           &  ${\rm sec}$     \\
   $\bar \theta$ & $15$& ${\rm deg}$       &$W_{tr}^p$         & $1.4 \times 10^{-3}$& *      \\
   $\theta_{lim}^t$ & $20$& ${\rm deg}$       &$W_{tr}^t$         & $2.0 \times 10^{-3}$& *      \\
   $ \mu_T$         & $0.05$&*          &$\epsilon_{SCP}^p$ & $10^{-5}$&    *   \\
   $ \tau_\theta$   & $1.0$ & ${\rm sec}$     &$\epsilon_{SCP}^t$ &$10^{-7}$ &  *    \\
   $\mathcal{F}^t$  &$4$    &${\rm Hz}$       &   $r_{lim}$                            & $10$&${\rm m}$ \\
\hline\hline
\end{tabular}
\end{table}
\begin{table}
\centering 
\caption{Tracking problem weights.}
\label{tbl:trackparam}
\begin{tabular}{ccc} 
\hline\hline
Parameter & Value & Unit  \\ 
\hline
   $\boldsymbol{W}_Q$      & $ 20 \cdot diag \left(1, 1, 3, \boldsymbol{0}_{1\times3}, 0.01, 0.01, 0 \right) $ &  *  \\
   $\boldsymbol{W}_R$      & $diag \left(1,1,0.1 \right) $ &  *  \\
   $\boldsymbol{W}_F$      & $     \begin{cases}
            \text{if } t + t_h < t_f^*  & 40*diag \left(1,1,3,1,1,3,0,0.1,0.1,0 \right) \\
            \text{else }  & 40*diag \left(0,0,5,5,5,15,0,0.1,0.1,0 \right)
        \end{cases}$ &  *  \\
\hline\hline
\end{tabular}
\end{table}
\begin{table}[!htb]
\centering
\caption{Initial conditions.}
\label{tbl:simIC}
\begin{tabular}{ccc} \hline\hline
\multirow{2}{*}{Parameter} & \multicolumn{2}{c}{Value}                              \\ \cline{2-3}
                           & IC 1                    & IC 2                     \\ \hline
$\boldsymbol{r}_0$                & $\left[-700, -700, -6000\right]^T\;{\rm m}$ & $\left[-780, -590, -6000\right]^T\;{\rm m}$  \\
$\boldsymbol{v}_0$                & \multicolumn{2}{c}{$\left[58.8, 58.8, 391\right]^T\;{\rm m/s}$}              \\\hline\hline
\end{tabular}
\end{table}

\subsection{Validation of Trajectory Planning}
To verify the accuracy of embedding atmospheric descent by the quadratic initial condition as given in Eq.~\eqref{eq:pinit} and linearizing the lift coefficient as given in Eq.~\eqref{eq:plift2}, a reference solution for Problem $\mathcal{P}^R$ is computed using a general-purpose optimal control software package (GPOPS-II)~\cite{GPOPS}. The problem solved by GPOPS-II is the same as Problem $\mathcal{P}^R$, except that atmospheric descent is included as an additional phase without approximation, and the expression of $\boldsymbol{F}_{lift}$ in Eq.~\eqref{eq:aero1} is used. For the purpose of comparison, compensated lift coefficients are not considered in this comparison.

\begin{figure}[!tb]
    \centering
    \begin{subfigure}[!tb]{7.5cm}
        \centering
        \includegraphics[width=7.5cm]{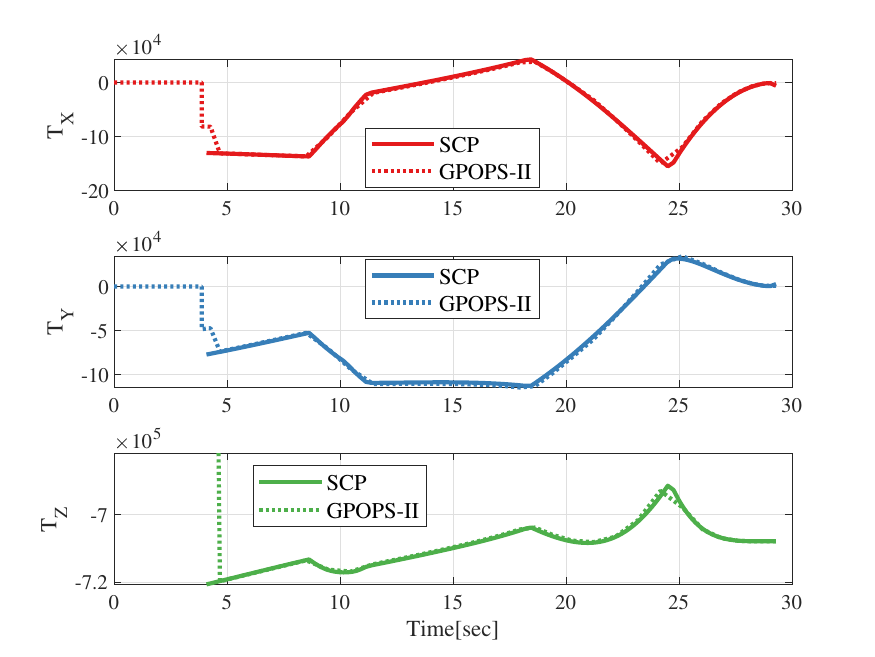}
        \caption{Thrust vector}
        \label{fig:thrustcomp}
    \end{subfigure}
    \begin{subfigure}[!tb]{7.5cm}
        \centering
        \includegraphics[width=7.5cm]{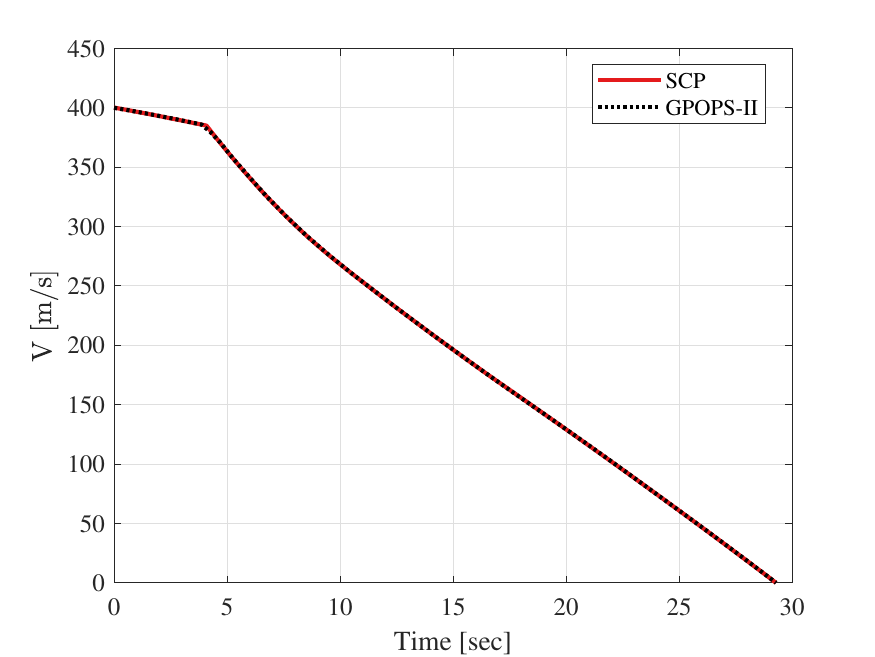}
        \caption{Velocity}
        \label{fig:velcomp}
    \end{subfigure}
    \begin{subfigure}[!tb]{7.5cm}
        \centering
        \includegraphics[width=7.5cm]{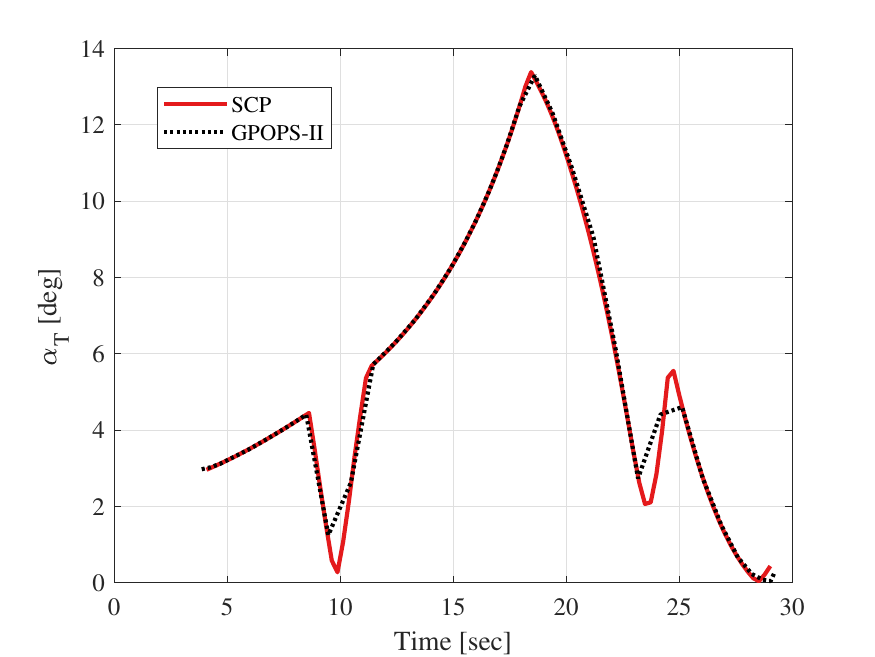}
        \caption{Total angle of attack}
        \label{fig:alptcomp}
    \end{subfigure}
    \caption{Comparison of optimal trajectory with GPOPS-II.}
    \label{fig:GPOPS}
\end{figure}

To thoroughly test the effect of approximation by linearizing the lift coefficient, IC 2 is chosen as the initial condition to illustrate the trajectory's similarity even with a relatively large angle of attack. The optimized trajectories are displayed in Fig.~\ref{fig:GPOPS}, revealing almost identical velocity and thrust profiles between the proposed method and GPOPS-II, with the same engine ignition time ($t_c^* = 3.88\;{\rm sec}$). Therefore, we can conclude that the accurate fuel-optimal trajectory is obtained by the proposed planning task. 

\subsection{Case Analysis}
\begin{table}[!tb]
\centering
\caption{Simulation cases.}
\label{tbl:cases}
\begin{tabular}{ccc} 
\hline\hline
Case & Initial Condition     & Algorithm                              \\ 
\hline
A1   & IC 1                  & Proposed  \\ \hline

A2    & \multirow{4}{*}{IC 2} & Proposed                               \\ 
B     &                       & Pure Drag                              \\ 
C     &                       & Uncompensated Lift                     \\ 
D    &                       & Analytic~\cite{Lu2023}  \\
\hline\hline
\end{tabular}
\end{table}
In this case study, we first test the baseline performance of the proposed guidance algorithm with an ideal initial condition as denoted by Case A1 in Table~\ref{tbl:cases}. Then, to qualitatively review the performance of the proposed guidance algorithm, we compare it with the three additional cases using IC 2. In Table~\ref{tbl:cases}, Case B is conducted with the proposed guidance, but the dynamics in both OCPs only consider drag, whereas Case C includes drag and lift force but does not employ the proposed lift compensation method. Lastly, to test the effectiveness of computational predictive tracking law, only the tracking task is replaced by the analytical tracking law proposed in~\cite{Lu2023}, where the thrust command is computed as follows:
\begin{equation} \label{eq:pinglu}
    \begin{gathered}
        \boldsymbol{T}_c     =   m  \cdot \left( \frac{\boldsymbol{T}^*}{m^*} - \frac{k_r}{h}\left( \boldsymbol{r} - \boldsymbol{r}^* \right) - \frac{k_r/2+1}{h^2}\left( \boldsymbol{v} - \boldsymbol{v}^* \right) \right)
    \end{gathered}
\end{equation}
Equation~\eqref{eq:pinglu} has two design parameters: $h$ and $k_r$, which are tuned to provide stable and sufficient performance ($h=5\;{\rm sec}$, $k_r = 4.836$). The thrust command is then converted to the attitudes and throttle command to match the direction and magnitude of $\boldsymbol{T}_c$.

\subsubsection{Baseline Performance}
\begin{figure}[!tb]
    \centering
    \begin{subfigure}[!tb]{7.5cm}
        \centering
        \includegraphics[width=7.5cm]{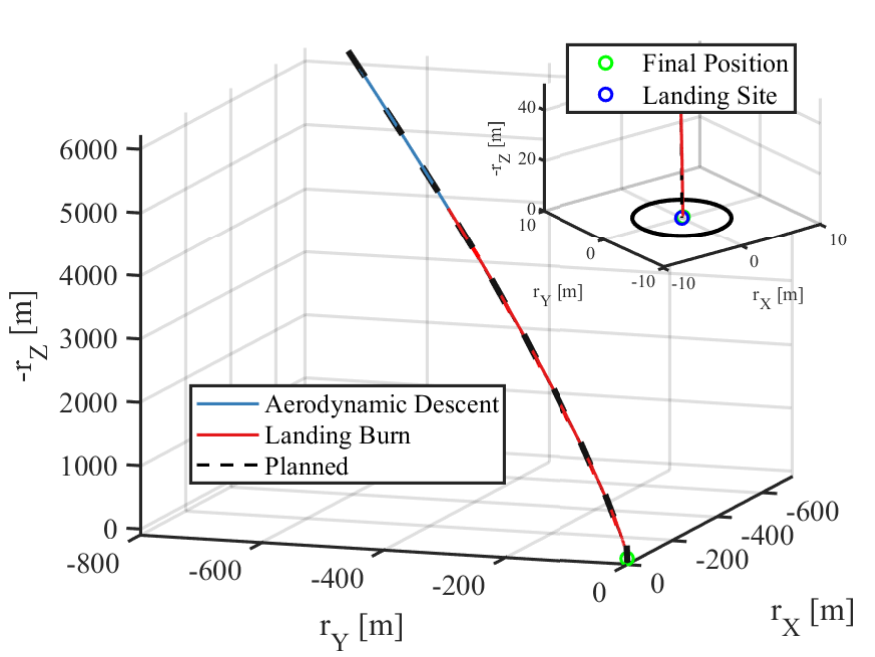}
        \caption{Trajectory}
        \label{fig:p3base}
    \end{subfigure}
    \begin{subfigure}[!tb]{7.5cm}
        \centering
        \includegraphics[width=7.5cm]{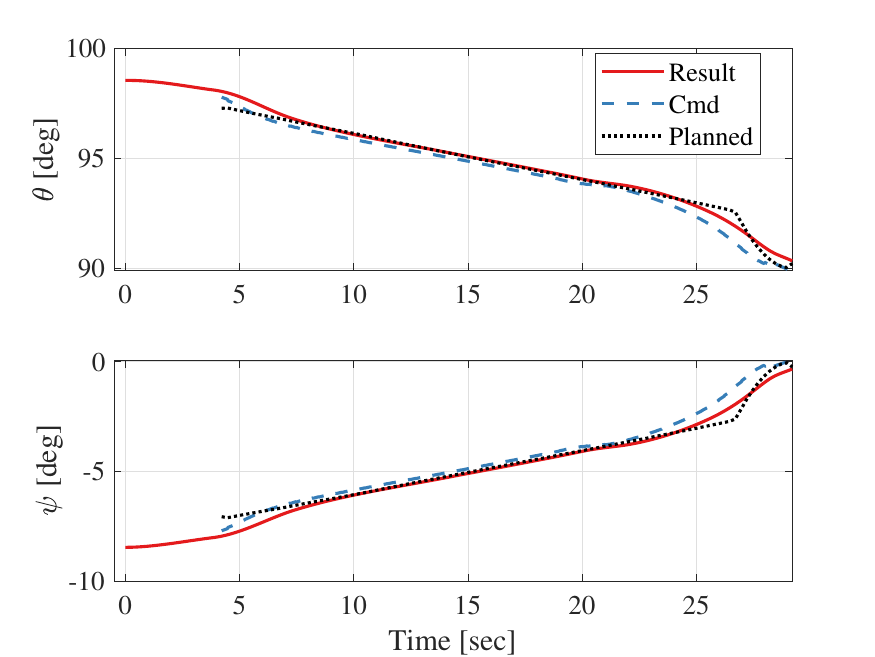}
        \caption{Euler angles}
        \label{fig:eabase}
    \end{subfigure}
    \begin{subfigure}[!tb]{7.5cm}
        \centering
        \includegraphics[width=7.5cm]{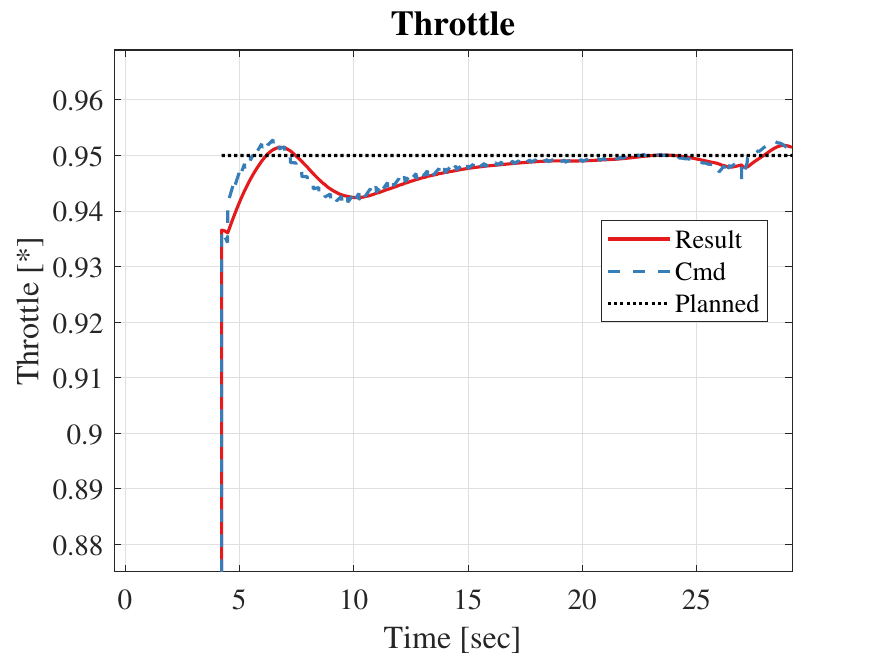}
        \caption{Throttle}
        \label{fig:throbase}
    \end{subfigure}
    \begin{subfigure}[!tb]{7.5cm}
        \centering
        \includegraphics[width=7.5cm]{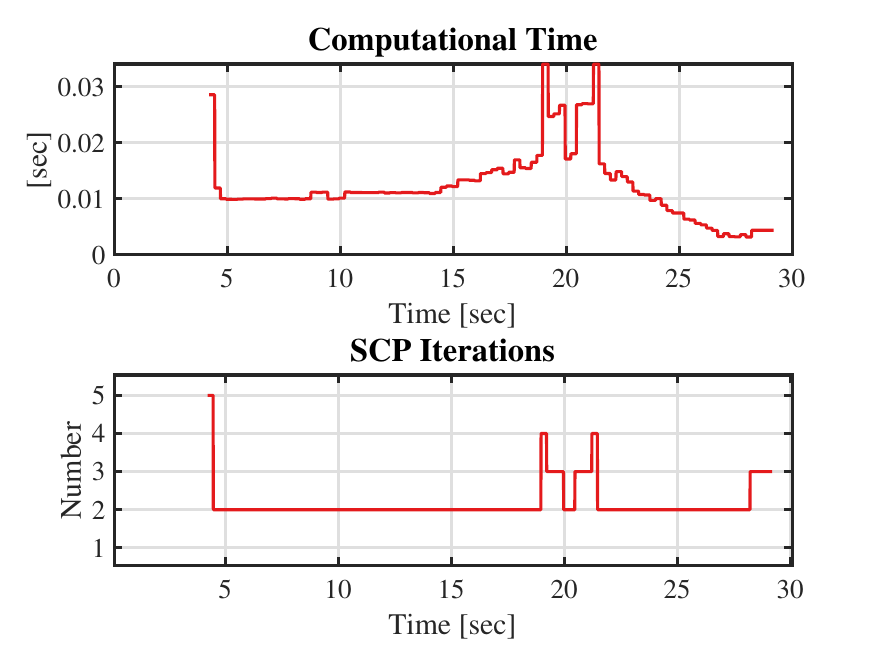}
        \caption{SCP of tracking}
        \label{fig:scpbase}
    \end{subfigure}
    \caption{Simulation result of Case A1.}
    \label{fig:caseA1}
\end{figure}
\begin{table}[!tb]
\centering
\caption{Final states of Case A1.}
\label{tbl:fsA1}
\begin{tabular}{ccccc} 
\hline\hline
$\lVert \boldsymbol{r}_f \rVert$ [m]& $\lVert \boldsymbol{v}_f \rVert$ [m/s]& $\theta_f$ [deg]& $\psi_f$ [deg]&  $\lVert \boldsymbol{\omega}_{B,f} \rVert$ [deg/s]\\ 
\hline
$0.320$ &$0.115$  & $90.35$ & $0.33$ & $ 0.567$  \\
\hline\hline
\end{tabular}
\end{table}
Figure~\ref{fig:caseA1} shows the simulation result of Case A1, and Table~\ref{tbl:fsA1} lists the values of final states. The throttle profile in Fig.~\ref{fig:throbase} is computed as $\lVert \boldsymbol{T} \rVert + A_e P_{atm}$ for visibility. The proposed guidance consistently tracked the planned fuel-optimal trajectory with minimal trajectory error, as seen in Fig.~\ref{fig:p3base}. Planning is conducted only once at the initial phase of the simulation. The SCP for Problem $\mathcal{P}_R$ converged to the optimal solution with only $5$ iterations in $0.215\;{\rm sec}$. The simulation is terminated at touchdown, with a horizontal position error of $0.320\;{\rm m}$ and a landing velocity of $0.115\;{\rm m/s}$. This demonstrates precise and soft landing performance, even though the design parameters in Table~\ref{tbl:trackparam} prioritize the final velocity rather than positional accuracy. Furthermore, due to the accurate tracking performance, the landing burn duration differs from $t_f^*$ by only a few milliseconds.

The guidance commands, comprising attitudes and throttle commands, display a continuous history in Figs.~\ref{fig:eabase} and~\ref{fig:throbase}, showing the ability to compensate for transient vehicle dynamics based on accurate attitude tracking performance in Fig.~\ref{fig:eabase}. Since the reference trajectory induces upright final attitudes with a thrust angle limit profile defined by Eq.~\eqref{eq:angacc}, the final attitudes are upright, and the magnitude of angular velocity is small as shown in Table~\ref{tbl:fsA1}. In terms of optimality, the initially planned fuel consumption is $7002.33\;{\rm kg}$, and the actual fuel consumption in Case A1 is $6980.83\;{\rm kg}$, which is $21.5\;{\rm kg}$ less than the planned value. This verifies the preservation of the fuel optimality by fine tracking performance. The reduced fuel consumption is potentially due to the relaxed thrust magnitude bounds in the tracking problem and slight trajectory deviation while tracking the reference trajectory.

The SCP iterations and computational time during the landing phase for Problem $\mathcal{T}$ are detailed in Fig.~\ref{fig:scpbase}. At the start of the landing burn, SCP requires $5$ iterations. The first initial guess for Problem $\mathcal{T}_2$ is generated by interpolating the reference trajectory, yet it remains considerably distant from the optimal tracking path. After the initial solution update, subsequent solution updates typically require $2$ iterations, with a maximum of $3$, as the preceding tracking solution offers a well-posed initial guess. The computational time peaks and averages at $34\;{\rm msec}$ and $12\;{\rm msec}$, respectively. The local peaks observed around $20\;{\rm sec}$ in the number of iterations and solving time result from a change in the cost function by weighting matrices in Table~\ref{tbl:trackparam} as the prediction horizon reaches the landing site. The actual computational time may increase when implemented in the avionics, potentially increasing by an order of magnitude. Nonetheless, given the roughly $20$-fold difference between the solution update period and the computation time, we conclude that the proposed guidance strategy holds substantial promise for practical implementations.

\subsubsection{Comparative Analysis}
\begin{figure}[tb]
    \centering
    \begin{subfigure}[!tb]{7.5cm}
        \centering
        \includegraphics[width=7.5cm]{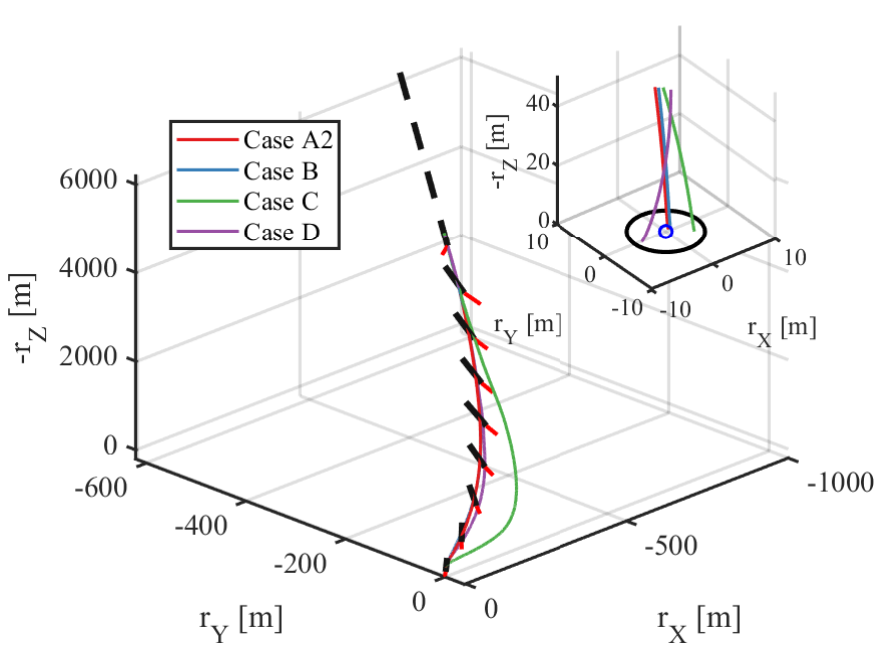}
        \caption{Trajectory}
        \label{fig:p3case}
    \end{subfigure}
    \begin{subfigure}[!tb]{7.5cm}
        \centering
        \includegraphics[width=7.5cm]{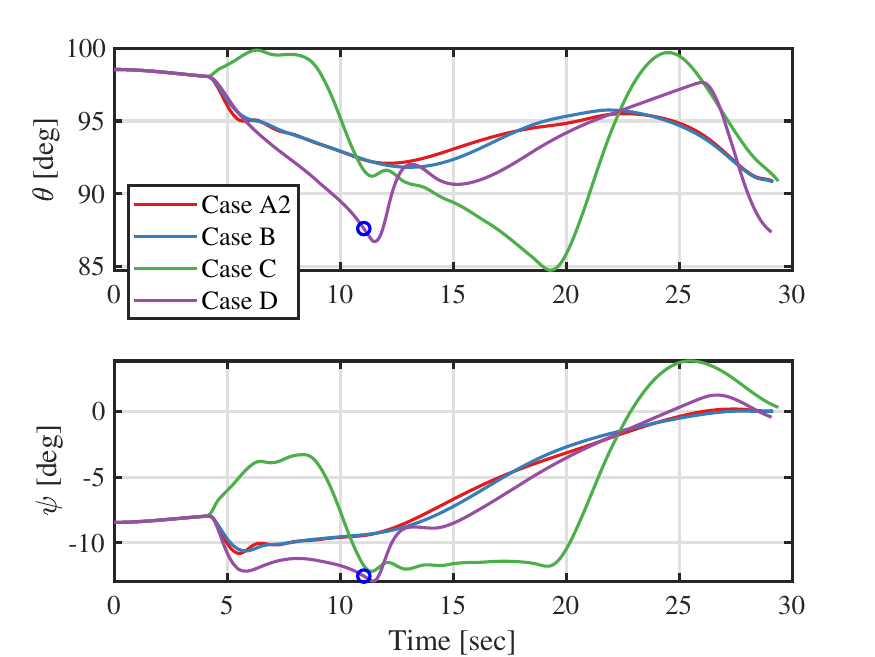}
        \caption{Euler angles}
        \label{fig:eacase}
    \end{subfigure}
    \begin{subfigure}[!tb]{7.5cm}
        \centering
        \includegraphics[width=7.5cm]{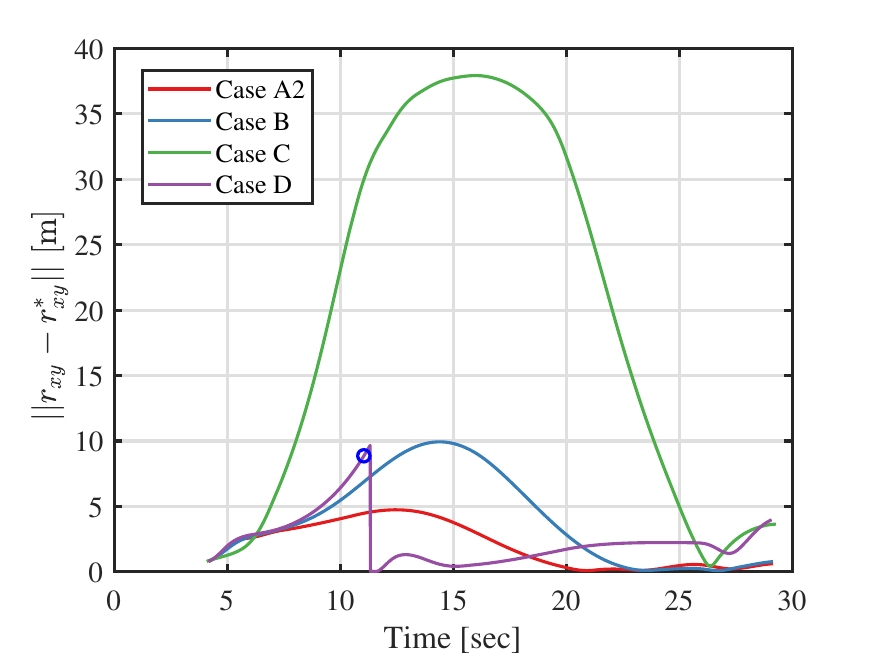}
        \caption{Horizontal position error}
        \label{fig:rerrcase}
    \end{subfigure}
    \begin{subfigure}[!tb]{7.5cm}
        \centering
        \includegraphics[width=7.5cm]{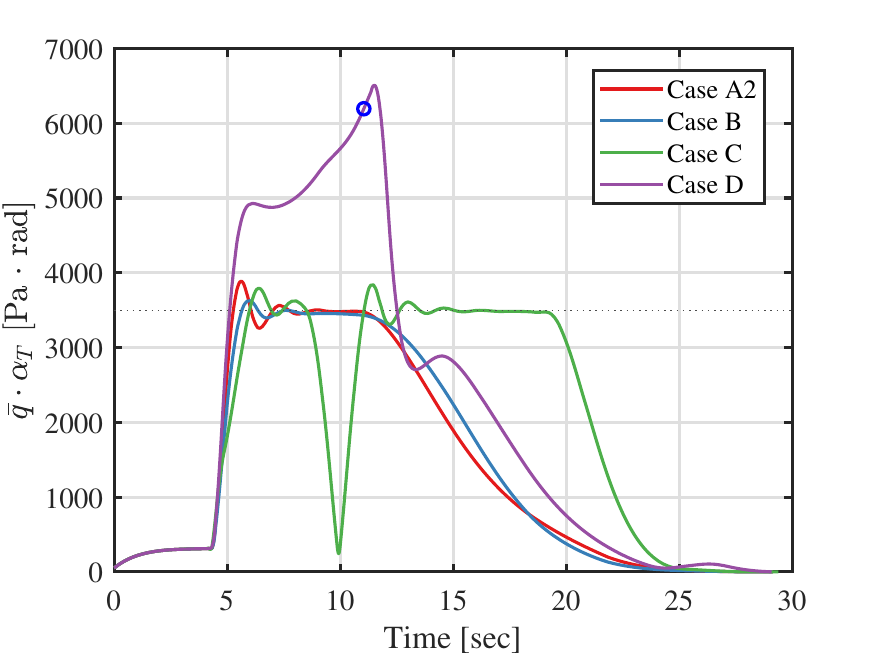}
        \caption{Aerodynamic load, $\bar q \cdot \alpha_T$ }
        \label{fig:aloadcase}
    \end{subfigure}
    \caption{Simulation results of Case A2, B, C, and D.}
    \label{fig:cases}
\end{figure}
\begin{table}[tb]
\centering
\caption{Final states of Case A2, B, C, and D.}
\label{tbl:fscases}
\begin{tabular}{cccccc} 
\hline\hline
Case &$\lVert \boldsymbol{r}_f \rVert$ [m]& $\lVert \boldsymbol{v}_f \rVert$ [m/s]& $\theta_f$ [deg]& $\psi_f$ [deg]&  $\lVert \boldsymbol{\omega}_{B,f} \rVert$ [deg/s] \\ 
\hline
Case A2&$\mathbf{0.621}$ &$\mathbf{0.460}$  & $90.832$ & $0.044$ & $ 0.920$  \\
Case B&$0.782$ &$0.536$  & $\mathbf{90.797} $ & $\mathbf{0.007}$ & $ \mathbf{0.852}$\\
Case C&$3.658$ &$1.124$  & $90.867 $ & $0.314 $ & $1.848$ \\
Case D&${3.987}$ &$2.476$  & $87.362$ & $ 0.430 $ & $1.101$\\
\hline\hline
\end{tabular}
\end{table}
Figure~\ref{fig:cases} shows simulation results for Case A2 through D, and the final states for each case are listed in Table~\ref{tbl:fscases}. Vehicle attitudes and the thrust vector of Case A2 are illustrated with landing trajectories for all cases in Fig.~\ref{fig:p3case}. The blue markers in Figs.~\ref{fig:eacase} through~\ref{fig:aloadcase} indicate the moment when the reference trajectory update is conducted. While all cases, except for Case A2, attempted trajectory re-planning when the total position tracking error exceeded $10\;{\rm m}$, only Case D successfully derived a new solution, with the remaining cases experiencing infeasiblity of Problem $\mathcal{P}_R$. 

Firstly, the proposed method (Case A2) exhibits the best performance in terms of position and velocity error at landing, as shown in Table~\ref{tbl:fscases}, with marginal differences in attitudes and angular rates from Case B. Also, the effectiveness of the proposed lift compensation method is evident in Fig.~\ref{fig:rerrcase}, as Case A2 shows the minimum value of position tracking error. Interestingly, Case C exhibits the poorest performance overall, despite having higher fidelity in terms of aerodynamics than Case B, which only considers drag force. This underscores the significance of the proposed lift compensation technique. The neglect of force deviation by the attitude control aspect of the system can lead to inaccurate leverage of lift force and eventually degrade the guidance performance, as observed in Case C.

Prior to the reference trajectory update, Case D has worse tracking performance than Case A2 and B, mainly because the tracking law in Eq.~\eqref{eq:pinglu} cannot consider aerodynamic forces. Nevertheless, Case D has a small amount of tracking error after successfully updating the reference trajectory. This implies that Case C can still have good tracking performance if the trajectory planning can be recursively conducted. However, having less dependency to the reference update through the proposed tracking law is still crucial, since solving fuel-optimal OCP is inherently susceptible to infeasiblity even with slight deviation during flight.

Moreover, Figure~\ref{fig:aloadcase} reveals that Case D violated the aerodynamic load constraint, as it cannot directly consider the complicated path constraints in derivation, whereas other trajectory optimization-based cases remained within the bound except for minor overshoot due to the attitude control loop. Lastly, Case D records the worst final position and velocity error, even though it has better tracking performance than Case C. This highlights the landing performance improvements by the model predictive methodology employed in the proposed algorithm. 

\subsection{Monte-Carlo Campaign}
\begin{table}[tb]
\centering
\caption{Dispersions for Monte-Carlo parameters.}
\label{tbl:uncparam}
\begin{tabular}{cccc} 
\hline\hline
Parameter & Range/SD & Unit & Distribution  \\ 
\hline
$\lVert \Delta \boldsymbol{r}_0 \rVert  $ &  100        & m      &  Normal  \\
$\lVert \Delta \boldsymbol{v}_0 \rVert  $ &  15         & m/s    &  Normal  \\
$ \Delta C_D  $                           &  [ -15, 15 ] &  \%    &  Uniform \\
$ \Delta C_L  $                           &  [ -15, 15 ]  &  \%    &  Uniform \\
$ V_w $                                   &  [ 0, 8 ]    &    m/s &  Uniform \\
$ \psi_w $                                &  [ 0, 360 ]  & deg    &  Uniform \\
\hline\hline
\end{tabular}
\end{table}
\begin{figure}[tb]
    \centering
    \begin{subfigure}[!tb]{7.5cm}
        \centering
        \includegraphics[width=7.5cm]{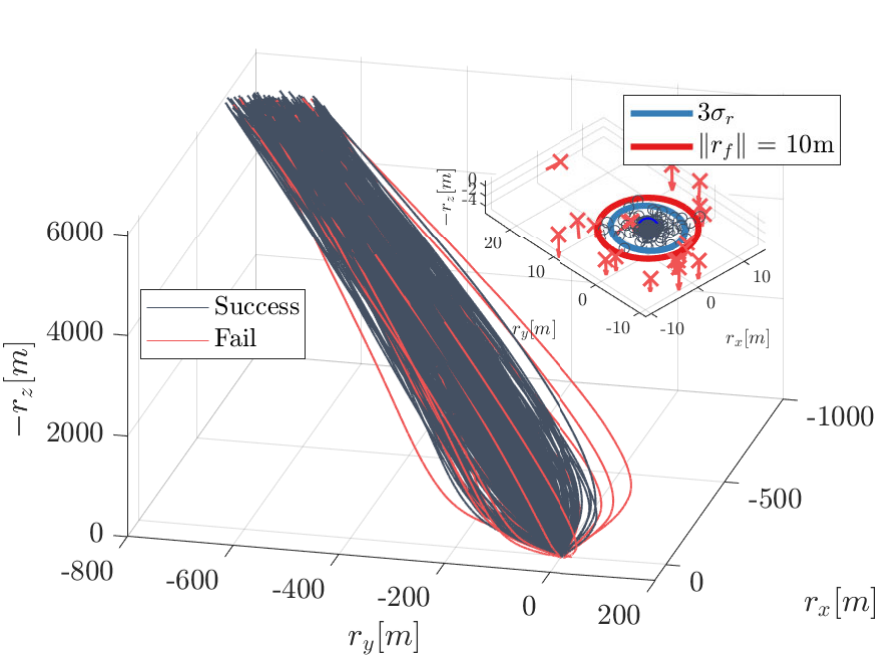}
        \caption{Trajectories}
        \label{fig:trajmc}
    \end{subfigure}
    \begin{subfigure}[!tb]{7.5cm}
        \centering
        \includegraphics[width=7.5cm]{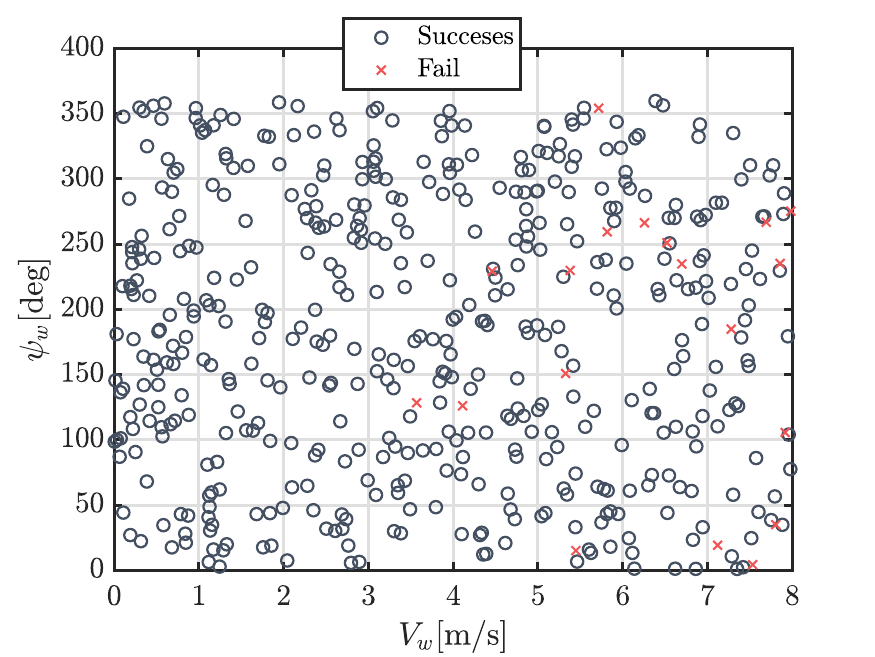}
        \caption{Wind uncertainties samples}
        \label{fig:scatterwind}
    \end{subfigure}
    \begin{subfigure}[!tb]{7.5cm}
        \centering
        \includegraphics[width=7.5cm]{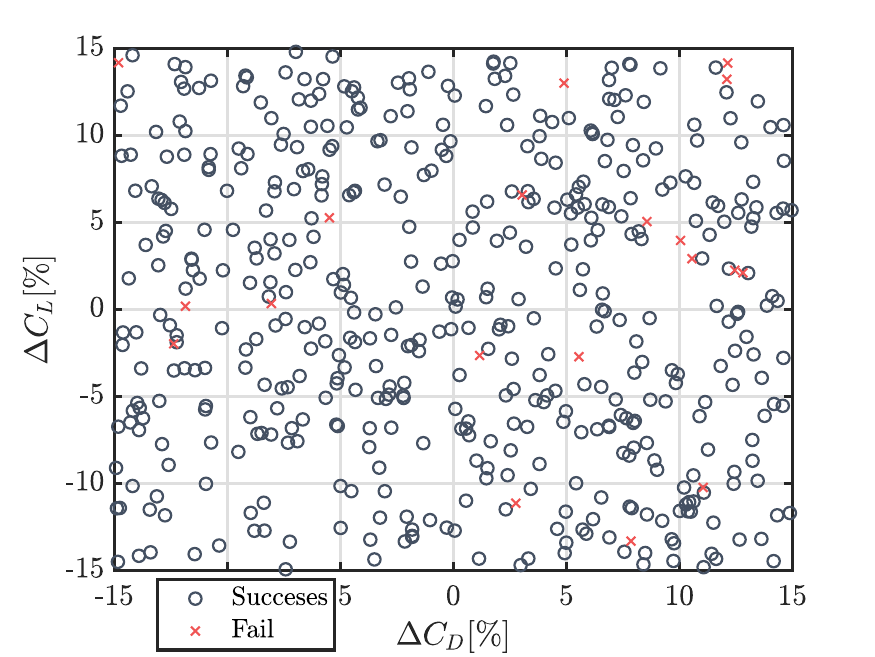}
        \caption{Aerodynamics uncertainties samples}
        \label{fig:scatteraero}
    \end{subfigure}
    \caption{Monte-Carlo simulation results.}
    \label{fig:MC1}
\end{figure}
\begin{figure}[!tb]
    \centering
    \includegraphics[width=7.5cm]{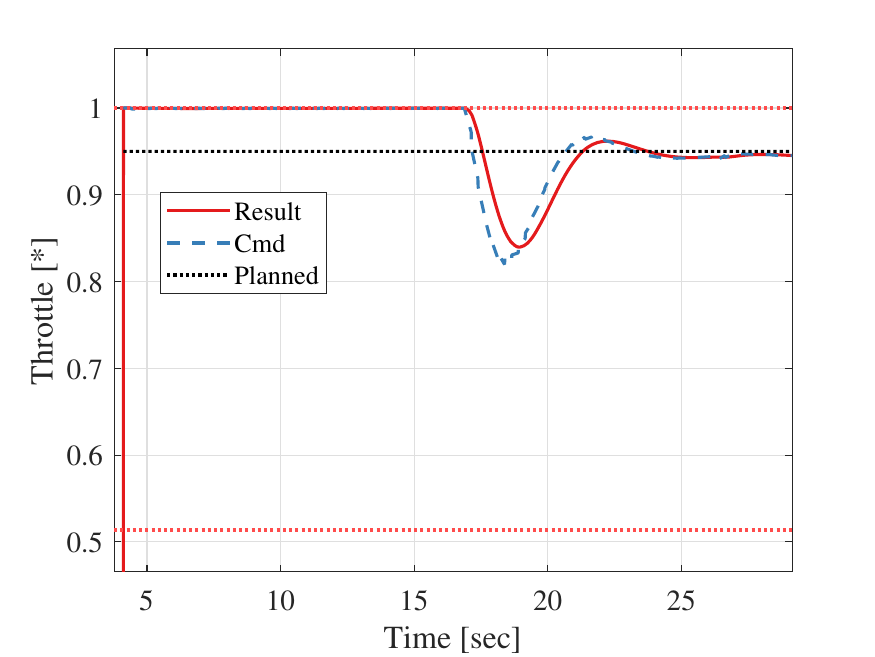}
    \caption{Throttle profile of sample \#54 ($\Delta C_D$ = -14.8\%, $\Delta C_L $= -14.5\%).}
    \label{fig:thro54}
\end{figure}
\begin{table}[tb]
\centering
\caption{Statistics of Monte-Carlo samples.}
\label{tbl:statMC}
\begin{tabular}{ccccc} 
\hline\hline
Statistics &$\lVert \boldsymbol{r}_f \rVert$ [m]& $\lVert \boldsymbol{v}_f \rVert$ [m/s]& $\theta_f$ [deg]& $\psi_f$ [deg] \\ 
\hline
Mean &$1.469$ &$0.693$  & $90.368$ & $-0.525$   \\
SD   &$2.402 $  &$0.719$  & $0.503$ & $0.804$   \\
\hline\hline
\end{tabular}
\end{table}
We test the performance and robustness of the proposed algorithm under various initial conditions and uncertainties through a Monte-Carlo (MC) campaign conducted with parameter dispersions in Table~\ref{tbl:uncparam}. The mean value of the initial condition is IC 1 and the directions of $\Delta \boldsymbol{r}_0$ and $\Delta \boldsymbol{v}_0$ are uniformly sampled. Considering the high-dynamic pressure environment, the aerodynamic models and wind are selected as the main sources of uncertainties. The guidance algorithm works with nominal values of $C_L$ and $C_D$, while dispersed coefficients $\tilde{C}_L = \left( 1 + \Delta C_L \right) C_L$ and $\tilde{C}_D = \left( 1 + \Delta C_D \right) C_D$ are used for the aerodynamics module in Fig.~\ref{fig:sim}. The variable $V_w$ is the wind magnitude, and the variable $\psi_w$ denotes the wind azimuth angle, with the assumption that the wind has only horizontal components.

Out of $500$ MC simulations conducted, $5$ samples that failed to acquire a reference trajectory in the initial phase are excluded from the subsequent analysis. All trajectories are drawn in Fig.~\ref{fig:trajmc}, and the arrows at the upper right corner indicate the scaled landing velocity. The scatter plots for wind and aerodynamics uncertainties in Figs.~\ref{fig:scatterwind} and~\ref{fig:scatteraero} show the influence of each uncertainty on successful landing. With the success criteria set as satisfying position error and velocity error under $10\;{\rm m}$ and $2.5\;{\rm m/s}$, $476$ samples successfully landed among the last $495$ samples, achieving the success rate of over $96\%$. The standard deviations of the final states in Table~\ref{tbl:statMC} remain reasonably contained, considering the failed samples also included in the calculation. The failed cases mostly show violations of both criteria, and the majority of the failures are due to the combination of severe aerodynamics uncertainties and wind, as failed samples are denser with larger magnitudes of wind and aerodynamics coefficient errors in Figs.~\ref{fig:scatterwind} and~\ref{fig:scatteraero}. Figure~\ref{fig:thro54} shows the throttle result of MC sample $54$, where severe aerodynamics uncertainties were sampled. By fully utilizing the thrust margin $\mu_T$, the proposed guidance achieved $\lVert \boldsymbol{r}_f \lVert$ and $\lVert \boldsymbol{v}_f\rVert$ of $0.633\;{\rm m}$ and $0.184\;{\rm m/s}$, respectively. In summary, the proposed landing guidance demonstrates fairly robust performance despite the presence of aerodynamics model errors and wind.

\begin{figure}[tb]
    \centering
    \begin{subfigure}[!tb]{7.5cm}
        \centering
        \includegraphics[width=7.5cm]{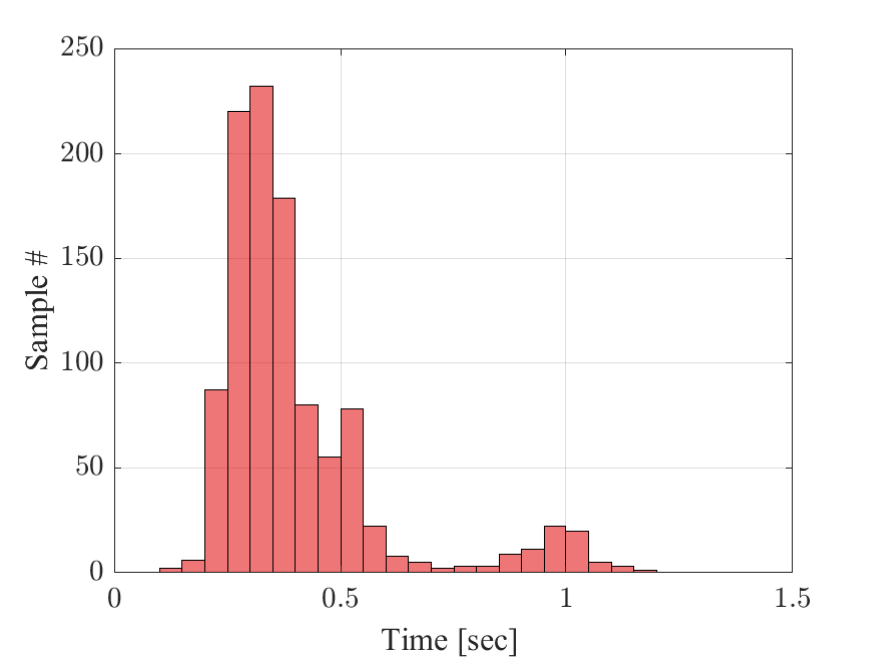}
        \caption{Computational time for planning task}
        \label{fig:histoplan}
    \end{subfigure}
    \begin{subfigure}[!tb]{7.5cm}
        \centering
        \includegraphics[width=7.5cm]{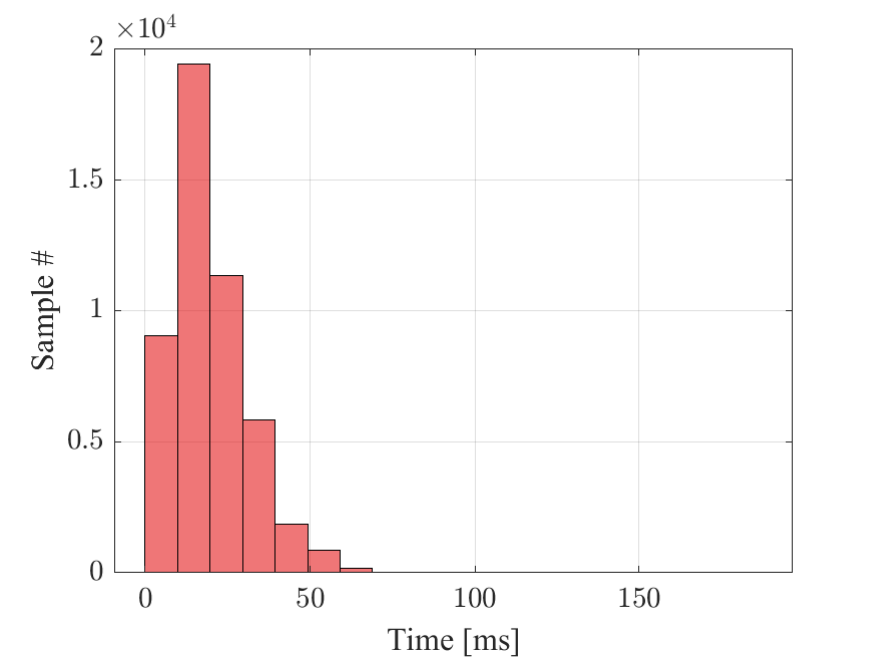}
        \caption{Computational time for tracking task}
        \label{fig:histotrack}
    \end{subfigure}
    \caption{Histogram of Monte-Carlo samples.}
    \label{fig:MC2}
\end{figure}

Lastly, the cumulative data of numerical optimization from all MC samples are presented as histograms in Figs.~\ref{fig:histoplan} and~\ref{fig:histotrack}. It is observed that solving Problem $\mathcal{P}^R$ and Problem $\mathcal{T}$ requires under $0.75\;{\rm sec}$ and $50\;{\rm msec}$, respectively, for the majority of MC samples. Additionally, although some cases experienced trajectory re-planning failures following the initial update, no samples exhibit guidance failures due to infeasible tracking tasks. Therefore, we conclude that the proposed algorithm demonstrates stable and fast numerical performance.

\section{Conclusion} \label{sec6}
In this paper, a fuel-optimal landing guidance strategy for a reusable launch vehicle (RLV) is presented, utilizing sequential convex programming for trajectory optimization. By separating trajectory planning and tracking into distinct optimal control problems, the proposed strategy significantly reduces computational demands, achieving guidance performance close to closed-loop control. Feasible states and controls are relaxed in tracking from planning, increasing the capability to overcome uncertainties and improve optimization stability. Modifications in the representation of aerodynamic forces, tailored for thrust-vector-controlled RLVs, effectively narrow the fidelity gap without adding computational complexity. The simplified representation of rotational dynamics in the tracking task enables optimal tracking commands that compensate for the transient nature of the RLV. Extensive 6-DOF simulation experiments validated the efficient computation by convex programming and the accurate landing performance of the proposed guidance. Through a Monte-Carlo campaign, the robustness of the proposed algorithm to model uncertainties was verified, along with stable numerical performance. The proposed algorithm demonstrated the benefits of applying trajectory optimization to both planning and tracking, improving precise and safe landing performance. These advancements promise to refine the precision and reliability of RLV landing operations.

\section*{Acknowledgments}
This work was supported by the National Research Foundation of Korea (NRF) grant funded by the Korea government(MSIT) (No. NRF-2022M1A3B8065627).

\bibliography{main}
\end{document}